# Electric field induced color switching
# in colloidal quantum dot molecules at room temperature.

Yonatan Ossia[1,2], Adar Levi[1,2], Yossef E. Panfil[1,2], Somnath Koley[1,2], Einav Scharf[1,2], Nadav Chefetz[1], Sergei Remennik[2], Atzmon Vakahi[2] & Uri Banin[1,2]*

[1]Institute of Chemistry, [2]The Center for Nanoscience and Nanotechnology, The Hebrew University of Jerusalem,

Jerusalem 91904, Israel

*Corresponding author: Prof. Uri Banin, uri.banin@mail.huji.ac.il

## Abstract

Colloidal semiconductor quantum dots are robust emitters implemented in numerous prototype and commercial optoelectronic devices. However, active fluorescence color tuning, achieved so far by electric-field induced Stark effect, has been limited to a small spectral range, and accompanied by intensity reduction due to the electron-hole charge separation effect. Utilizing quantum dot molecules that manifest two coupled emission centers, we present a novel electric-field induced instantaneous color switching effect. Reversible emission color switching without intensity loss is achieved on a single particle level, as corroborated by correlated electron microscopy imaging. Simulations establish that this is due to the electron wavefunction toggling between the two centers dictated by the electric-field and affected by the coupling strength. The quantum dot molecules manifesting two coupled emission centers may be tailored to emit distinct colors, opening the path for sensitive field sensing and color switchable devices such as a novel pixel design for displays or an electric field color tunable single photon source.

## Main

Semiconductor quantum dots (QDs), prepared either by molecular beam epitaxy or via colloidal wet-chemical synthesis, have reached the breakpoint of commercial applications in technologies[1–3] ranging from biological tagging[4–6], displays[7], lasers[8,9], and as single photon sources for quantum technologies[10–12]. Optical tunability is key for these applications and is currently realized chiefly at the QD's growth stage. Post-synthesis color tunability is desired for novel multifunctional devices including a tunable QD laser, tunable QD pixel display, and tuning the emission in single and entangled photon sources for quantum interference devices[13]. A most promising approach towards such tuning has been via the application of external electric fields inducing a Quantum confined Stark effect.[14–17] The bands are bent by the external potential dynamically altering the excitonic energy and thus enabling potential functionality as a tunable light source. However, the main effect is the manipulation of electron-hole charge separation in the presence of the field, which limits the color tuning range while repressing the radiative recombination pathway leading to decreased intensity[18–20]. Essentially, this limits their electric field (EF) response for sensing applications to emission intensity modulation.

Herein, we introduce a novel pathway for electric-field induced broadband color switching realized on coupled colloidal quantum dot molecules (CQDMs)[21]. Such systems manifest two emission centers, with photoluminescence (PL) energies that are separately dictated by their sizes and composition. Their fusion forming a CQDM, leads to delocalization of the electronic wavefunctions and enables instantaneous color switching between the two centers upon application of the EF without significant overall intensity modulation, and under ambient conditions. Prior reports on epitaxial "quantum dot molecules" reveal coupling effects observable only at low temperatures, limiting their applicability.[22–25] The application of EF was studied, revealing mainly avoided crossing effects upon reaching the resonance condition between the two composing dots.[26] Therefore, the primary effect of applying an electric field was a transfer from a spatially direct to a spatially indirect exciton transition as the electron and hole are pulled away from each other under the field.[27] The fluorescence is thus typically tunable by merely a few meV, resolvable only at cryogenic conditions, and this color shifting is usually accompanied by intensity modulation.

In the colloidal quantum dot molecules studied herein, the chemical flexibility in choosing the material system and its small dimensions opens a different regime of EF induced color switching. Utilizing CdSe/CdS core/shell quantum dot building blocks as a model system, the resultant CQDM manifests a "quasi-type II" character[28]. Unlike the hole, the electron is easily toggled between the two emission centers under the electric field, due to its substantially lighter effective mass and the lower potential energy barrier in the conduction band (CB). Furthermore, in the CQDM, the coupling strength is an order of magnitude larger[21,28] than in epitaxial QDMs,[22,29,30], and is tunable by controlling the neck region connecting between the two emission centers[31,32]. Altogether, this enables a novel behavior of actual color switching between the two centers, that provide substantially distinct emission colors. Considering also the inherent flexibility in incorporating nanocrystals in diverse media via bottom-up fabrication, this opens the path to a new family of nanoscale color-tunable emissive devices controlled by an electric field, for applications ranging from voltage sensing in neuroscience, to color-switchable pixels in displays and for chip-scale color switchable quantum emitter devices operable up to room temperature.

**Results and discussion**

The EF induced color switching is demonstrated on a model CQDM system consisting of two fused CdSe/CdS core/shell QDs prepared via dimer formation on a template of silica nanospheres[21,31], followed by fusion at moderate temperatures forming a connecting CdS lattice (Fig. 1A, Supplementary Fig. 1). The EDS elemental mapping clearly shows the two CdSe cores overcoated and linked by the CdS shell. In this model system, the color difference between the two emission centers ranges up to 100 meV related to variations in the mono-QD's core/shell size distribution (Supplementary Fig. 2c).

For the electric field induced switching experiments on single CQDMs, a dilute solution was spin cast onto an in-house fabricated interpenetrating electrode device printed on a glass coverslip (see methods). The external EF between the electrodes is periodically modulated on-off while exciting at 450 nm, well-above the band gap, and collecting the single particle PL spectrum, polarization, time-tagged-time-resolved (TTTR) photon statistics, and second-order photon coincident counts (referred as $G_2(\tau)$) in a synchronized manner (Supplementary Fig. 3). Figure 1B, C present two PL spectra manifesting typical behaviors of the single particles with and without EF assigned to single or dual emission centers, respectively. The single emission center case shows negligible change upon applying a 180V potential field, while the dual emission center PL shows a color switching effect manifested as a blue shift of 27±2 meV. Its behavior is well represented by a combination of two peaks, "red" and "yellow", spaced by 51±8 meV. Without field, the contribution of the red and yellow peaks to the PL is nearly equal, while upon applying the field, the emission is nearly fully switched to the yellow center. Furthermore, the measured emission polarization characteristics manifest elliptical polarization with the angle of its principle axis relative to the electrodes changing between emission centers (177°±5° and 94°±5 for the red/yellow cores of the CQDM, respectively, Supplementary Fig. 4). As for spherical wurtzite QDs the emission polarization is perpendicular to the C-axis,[33,34] This is also indicative of heteronymous plane attachment in the CQDM, in which the two QDs attach and fuse through different planes[31,35].

To further prove the hypothesis of the different PL behavior being attributed to a mono-QD versus a dual center CQDM, we introduce a novel method to directly correlate the emission of single nanocrystals (NCs) with high resolution STEM imaging (see methods, Supplementary Fig. 5 & movie M1). Figure 1E presents the PL image showing bright single particle spots in between the ruler-marked electrodes. The same region was imaged by scanning-electron microscopy (SEM) within a dual SEM-FIB (focused ion-beam) microscope (Fig. 1F). A lamella of the region containing the two PL spots discussed above, is then cut out by the FIB, and placed onto a TEM grid. Fig. 1G shows the HAADF-STEM image revealing the presence of two single particles, a mono-QD in the top left side, and a CQDM in the bottom right, which directly correlates with the locations of the PL spots. This unequivocally establishes that the single-center emission case is associated with a mono-QD while the dual center emitter is a CQDM. The correlation also enables to directly extract the orientation angle of the CQDM axis, relative to the electric field (45°) and we can infer that the yellow emission center is the upper dot, considering the field polarity.

Upon establishing that the color switching effect is related to the CQDM, we further examine this phenomenon at 10 Hz EF modulation rate (Fig. 2). The field modulation results in instantaneous color switching of the PL spectra (Fig. 2A-C, Movie M2), yet does not alter the emission intensity also as measured by an Avalanche photodiode (APD). The corresponding time resolved emission traces are well fitted to a bi-exponential decay (Fig. 2D, upper box, Supplementary Fig 6). The fast, ~1.25ns component, is mostly unresponsive to the applied field and is assigned to a fast-non-radiative Auger decay process of either charged or bi-exciton states[21,32]. The slow component, reflecting radiative decay of the neutral exciton, is slightly shortened (elongated) under the positive (negative, a sequential measurement) bias consistent with a dominant PL contribution from the smaller-upper- "yellow" (bigger-bottom- "red") QD emission center possibly due to volume[36] or trap profile differences[37,38] between emission centers under the applied EF. Single QDs and CQDMs often show in-measurement emission spectral diffusion and blinking[39–41] due to charging and surface trapping. Fig. 2E presents gray bars showing the CQDM spectral shifts between subsequent "without EF" frames where the mean frame shift is 0 meV (Fig. 2F, gray). On the other hand, the energy shifts between subsequent EF modulated frames of 180V-0V (red) exhibit a mean frame shift of +27±1 meV switching to higher energy emission throughout the entire measurement, with a normally distributed probability resulting from the spectral diffusion. A sequential measurement, applying an opposite polarity EF, results in emission shifts to lower energies related to the red core, in correspondence with the color switching phenomenon (Fig. 2D,E,F, blue, see Supplementary Fig. 9 for other voltages).

The EF induced color switching effect is unique to CQDMs with two emitting centers, while field induced PL shifts in Mono-QDs mostly result from modulation of the electron-hole charge separation,[16,42] which at the same time reduces the emission intensity depending on their overlap. For the CQDMs, the field can alter the exciton populations between the cores leading to color switching without changing the emission intensity. To understand the color-switching mechanism, we simulated the excitonic emitting states for the CQDM by solving the self-consistent Schrödinger-Poisson equations (using COMSOL Multiphysics) following our previously reported algorithm[28,35]. The electron and hole wave-functions and energies were calculated, upon adding an external electric potential[16] (methods, Supplementary Fig. 10) that creates a field between ±500 kV/cm (similar to the voltage induced fields used in the experiment). The dimensions of the CQDM's outer shell were taken from the HR-STEM image, while the core sizes were taken to fit the peaks of the accumulated spectra (in Fig 1G), resulting in a core size difference of 3.5Å, indicating one additional CdSe monolayer[43] in the red emitting center. The hole in this system is characterized by a heavy effective mass and large core/shell band offset, leading to its localization and confinement to either of the QD cores even at high EFs (Supplementary Fig. 11). Following excitation and relaxation to the band edge state, the hole branches to either the red or yellow core. The electron on the other hand, is characterized by a light effective mass and a low core/shell band-offset and thus can be easily delocalized across the CQDM under external applied EF. Without EF, the electron-hole are attracted by coulomb interaction to overlap in the same emission center, forming the direct excitons (Supplementary Fig. 12). Simulating the electron states, with the hole in a specific side QD-core localized state, we show that upon applying the field in a direction opposite to the hole-localized core, a sufficient applied field potential can overpower the attractive electron-hole coulomb interaction and switch the electron wavefunction to the other side of the CQDM preventing emission from the center with the initially localized hole (Supplementary Fig. 13). This effect is represented by the EF dependent electron-hole wavefunction overlap that its square magnitude is proportional to the radiative rate[6,16,19], normalized to the overlap at zero field (Fig. 3A/B for the hole in the left/right emission centers, respectively):

$$\frac{\langle \Psi_e^* | \Psi_h \rangle_{\vec{E}}}{\langle \Psi_e^* | \Psi_h \rangle_0}$$

For a field direction parallel to the dimer axis (Fig. 3, red), the EF switching threshold is at ±125 kV/cm, above which the electron switches over to the other core. At a field perpendicular to the dimer axis, (Fig 3, gray), the coulomb interaction between the charges dominates and dictates the placing of the electron, regardless of the field magnitude. The calculated emission energy shift with applied EF shows only small changes with a quadratic dependence due to the expected field induced charge separation on the same QD (1-9 meV, Supplementary Fig. 11). At 45° (Fig. 3, yellow), color switching also takes place at ~250 kV/cm, with a slightly reduced efficiency. The simulation thus explains the experimentally observed color switching behavior in the CQDMs as due to EF induced electron toggling between red/yellow emission centers. The hole itself is not directly toggled by the EF, and beyond its statistical decay to either emitting center, it may also be affected by the coulomb interaction with the electron favoring its relaxation to the same side and forming the spatially direct exciton. This explains the observed EF induced switching of the red/yellow emission ratio, while the overall PL intensity is nearly unchanged. This color switching behavior differs significantly from the EF induced charge separation effect observed so far for nano-systems with a single emission center[16,18,44].

Addressing the generality of the EF induced color switching effect and its dependence on various parameters, Fig. 4 presents additional data for three CQDMs. Shown are the area-normalized accumulated spectra under different fields (A-C), the related mean-frame energy shift analysis (D-F), and the average area ratio between the "higher" to "lower" energy peaks normalized to the area ratio at 0V (G-I) as an additional measure of the EF switching efficiency separating out intrinsic emission yield fluctuations. Fig. 4J-L show the normalized second order photon correlation coincidence - $G^2(\tau)$, from which the area ratio $g^2(0)$ is extracted: $\frac{\int G2_{0-pulse}}{\int G2_{next\ pulse}} \approx \frac{BX_{QY}}{X_{QY}}$ where $G^2(pulse)$ is the fitted second order correlation function that was shown for monomers to be equal to the biexciton-exciton emission quantum yield ratio at $\langle N \rangle \ll 1$.[45] Low $g^2(0)$ characterizes a single photon emitter, while high value indicates a multi photon emitter. Our recent work on the photon statistics of CQDMs[31,32] indicates that there is a correlation amid the filling of the neck between the two fused QDs and the value of $g^2(0)$. At weak neck filling, the weak coupling leads to a behavior closer to two separate emitters leading to a rise of $g^2(0)$ by segregated biexciton emission. Upon filling of the neck, the electronic coupling increases and the energy states are more delocalized, enhancing multi-exciton non-radiative recombination rates due to enhanced Auger interaction and leading to reduced $g^2(0)$ [46–48].

We find that strongly fused, single photon emitting CQDMs can show a complete turning off of one of the emission centers (Fig 4a, +180V), for at least one direction of the field polarity. This color switching is also noticeable in the mean frame shift analysis where the +180V to 0V frame shifts (red) are larger than the shifts originating from spectral diffusion. The second CQDM, with higher $g^2(0)$ correlated with weaker coupling (Fig. 4b), is an example showing the presence of two spectrally separated emission centers but exhibiting less complete color switching under the criteria of the change in the PL "yellow"/"red"-peak area ratio. Interestingly, the $g^2(0)$ also changes somewhat under the field, between values of 0.37-0.28-0.24 (Fig 4k) for +180V, 0V, -180V EF modulation, in parallel with the reduction of emission from the "yellow" emission center, indicating possible segregated biexcitons where each photon emission arises from the different cores. A special case of a weakly fused hetero-CQDM with two clearly distinguishable emission peaks (Fig 4c) shows a lower "yellow"/"red" modulation ratio under the criteria of the area ratio. At the -180V state, the EF mostly "switches-off" the red center, and apparently the emission intensities at both colors are nearly equal at certain times (Supplementary Fig 14). This leads to large differences in the fitted frame-by-frame emission energy shifts (Fig 4f, gray/blue).

A fuller representation of the color switching effect is provided in Fig. 4M, showing the dependence between the mean frame spectral shifts to the measured $g^2(0)$ for 89 particles, where stars mark cases for which structural correlation with STEM was also performed (Fig. 4-O, extended details in Supplementary Fig. 15-19).

A general correlation between low $g^2(0)$ (strong inter-particle fusion) to high spectral shifts is observed in CQDMs. Mono-QDs from the sample prior to CQDM formation, characterized similarly, show only small spectral shifts (average shift - 2.5meV, Supplementary Fig. 2f). They also manifest low $g^2(0)$ and nearly mono-exponential lifetime decay for the high intensity counts of the bright emitting on-state (Supplementary Fig. 7-8). CQDMs show a multi-exponential decay even for the bright tags (Supplementary Fig. 6,16-19). Thus, mono-QDs in the CQDM sample are identified by a combination of their lifetime characteristics, as well as STEM correlations (Fig. 4M, yellow squares). CQDMs from this sample, with a $g^2(0) < 0.1$ (single photon emitter regime) show color switching with applied EFs, with mean energy shifts significantly exceeding maximal observed mono-QD shifts by more than 2-fold. Higher $g^2(0)$, indicating a lower degree of inter-particle fusion, correlates with lower EF induced modulation, with the very weak to non-fused regime (value above 0.4) showing mono-QD like small EF induced shifts. This indicates that a key characteristic affecting the color switching extent is the inter-particle coupling.

To further establish the importance of facile coupling for the EF color switching, we simulate the effect of the potential barrier characteristics for the same CQDM geometry as before, showing the EF effect on an exciton forming in the yellow emission center. First, probing variations of the CB offset (Fig. 5a, Supplementary Fig. 20) for a range of literature values between 0-0.3 eV,[28,49,50] we find that while the 0-0.1 eV CB offsets show electron switching for EFs higher than 125 kV/cm, the 0.2-0.3 eV offsets have metastable "S" wavefunction like states in both cores also for 250 kV/cm, and color switching may only occur at fields higher than 250 (500) kV/cm for the CB offset of 0.2 (0.3) eV, respectively. Second, for the 0.1 eV offset case, we examined the effect of the inter-particle neck filling (Fig 5, B) ranging from "fully fused" (7.5 nm width), to "touching" (0.6 nm width). Here, the electron wavefunctions are strongly affected, and for each different neck width, there is a different threshold EF over which electron switching is predicted to the core opposite of the applied field direction. For the "touching" state, the energy levels of the electron are localized in each of the emitting centers and the barrier prevents cross-particle electron relaxation between the spatially segregated states thus preventing the EF color switching. Upon neck filling, the barrier between the fused QDs is reduced resulting in delocalization of the electron states and allowing facile spatial modulation of the electron by the EF (Supplementary Fig. 21). This enables cross-QD electron relaxation at gradually lower field thresholds: 375 kV/cm for a 3 nm neck, 250 kV/cm for 5.3 nm, and 125 kV/cm for a "full" neck. At 125 kV/cm though, the first (indirect) and second (direct) exciton states are spaced by mere 13 meV for the full neck and up to 25 meV for the 3nm neck, meaning that at room temperature thermal occupation may play a role as well, counteracting the color switching effect.

Fig. 5C-E, illustrate the findings. Considering the situation of the hole residing in the higher energy emission center without external EF, coulomb interaction between the charge carriers is dominant inducing exciton localization on that side. The barrier for the hole between the two cores prevents inter-particle relaxation and emission from that center is "on". Applying the field, the indirect exciton state is lower in energy despite coulomb interaction. However, the actual switching-"off" depends on the barrier between the two centers. While in case of strong coupling, the electron will easily switch over to the indirect exciton state turning-off that emission channel, in weak coupling it may remain in the metastable direct exciton state and emission persists.

Additionally, we simulated the effect of core location offset from the QD center (as can be seen in some of the cases in Fig. 1-A) on the color switching threshold (see supplementary figures 22,23). The simulation shows that if the core location is closer to the second QD, the electron delocalization is enhanced at low EFs. High EFs that should toggle the electron to the second QD show the same switching phenomenon as a centered core, along with a reduced (increased) electron-hole wavefunction overlap of 67% (110%) compared to a centered core at that side, due to a core location offset of 2nm towards (opposite) the direction of the second QD. The combination of this effect with weak intra-particle fusion can explain some of the CQDM measurements that showed intensity loss as well as color switching (Fig. 4-C).

To further demonstrate the potential color range of this approach, we simulated the switching effect for a hetero-CQDM comprised of an extremely small "green" core emitting at 542nm and a large red center emitting at 624nm (Fig. S24). The switching effect is also achieved, albeit at somewhat larger bias for uphill switching towards green emission. This prediction is supported by experimental data. To demonstrate the feasibility for broadband color switching, we created two distinct samples of CQDMs synthesized from monomers with a broad size distribution (supplementary Fig. S25). In figure 6-A-B we show an example of a Hetero-CQDM made from large (~15nm & ~12nm diameter) QDs, showing broadband color switching between "red" (647nm) and "yellow" (600nm) emitting cores. The 10Hz EF modulation shows robust switching throughout the entire measurement, with an average PL shift of 150 meV (~50nm) upon applying opposite field polarities (supplementary figure S26-27, movie M3). Fig. 6, C-D shows an example of color switching between "yellow" (590nm) and "green" (555nm) emitting centers in a Hetero-CQDM made from small (~9nm & ~6nm diameter) QDs. The PL color switching is seen mainly in the accumulated spectra and peak ratios, and less clearly in the 100ms frame by frame analysis (supplementary figures S28-30). For these small sizes, the initial QDs exhibit less stable emission and spectral blinking. Due to the thinner shell, they are more prone to have structural defects and formation of a continuous lattice between coupled QD is less probable. We note that for the second sample we achieved significant EF affect when using devices with 2.5 μm separated electrodes, and a Polyvinylpyrrolidone (PVP) as the filler polymer. These changes enhance the electric field between the electrodes (+25% compared to 3.1 μm spaced electrodes) and reduce the dielectric screening between the CQDM and surroundings ($\varepsilon_r(PVP) \approx 7.7, \varepsilon_r(PMMA) \approx 2 - 4$). The combination of these modifications in the sample preparation lead to a significantly stronger effect of applied EF on the charge carriers. For ±90V, color switching without PL intensity loss is seen. The high ±180V applied EFs, additionally increase the electron-hole charge separation in both CQDM emission centers, which reduces the radiative recombination rates and yields lower PL intensity accompanied by energy shifting to lower energies, as is commonly seen for excitons in mono-QDs under a strong EF Stark effect (supplementary Fig. S31).

**Conclusions and outlook**

Facile EF induced color switching at room temperature is shown in strongly fused CQDMs uniquely manifesting two emitting centers that can be separately controlled via tailoring the size and composition of the system. Direct STEM structural correlation with single particle emission, as well as numerical simulations, confirm that the color switching effect is associated with induced electron toggling between two coupled emission centers. Additionally, these correlations strengthen previous works that linked structural heterogeneity and in-plane orientation to optical properties at the single particle level,[51–54] while our novel method using a FIB lift-out process can be utilized in a variety of configurations and materials as well as operating chip-scale devices.

Our current EF modulated CQDM devices can serve as color switching single photon sources operating in both cryogenic and room temperature, while already showing record high 50nm color switching, higher than any other present-day active color switching nano-device (to the extent of our knowledge). We emphasize that the extent of this effect can further be increased if making CQDMs from QDs with different core materials or a type-II band alignment and achieving better chemical fusion between dots. Creating bright color switching devices from an array of CQDMs requires control on ensemble properties of the CQDMs, such as deposition in parallel to the applied EF direction of multiple particles, and use of a monodispersed solution with high CQDM to mono-QD ratios. Aligned NC deposition between electrodes has been shown by deposition under an AC applied EF possibly combined with photoexcitation[16,55], where the rotation of colloidal NCs is driven by the excited-state dipole moment in solution and additionally implemented to solid state films.

This study therefore lays the ground for potential applications of the color switching effect. For example, for a new type of color-switchable red-green display pixel that potentially can simplify the classic RGB display production, or for an electric field color tuned single photon source. There is also a potential to apply the effect towards a novel approach for sensing electric fields in neuroscience.

## Acknowledgements

The research leading to these results has received financial support from the European Research Council (ERC) under the European Union's Horizon 2020 research and innovation programme (project CoupledNC, grant agreement No [741767], and project CQDplay, grant agreement No [101069322]). Y.E.P. acknowledges support by the Ministry of Science and Technology & the National Foundation for Applied and Engineering Sciences, Israel. S.K. acknowledges the support from the Planning and Budgeting Committee of the higher board of education in Israel through a fellowship. U.B. thanks the Alfred & Erica Larisch memorial chair. We thank Dr. Shimon Eliav, Dr. Itzhak Shweki, Mrs. Galina Chechelinsky and Mr. Maurice Saidian from the unit for nanocharacterization (UNC) of the Hebrew University Center for Nanoscience and Nanotechnology for assistance in the electrode device fabrication.

## Figures:

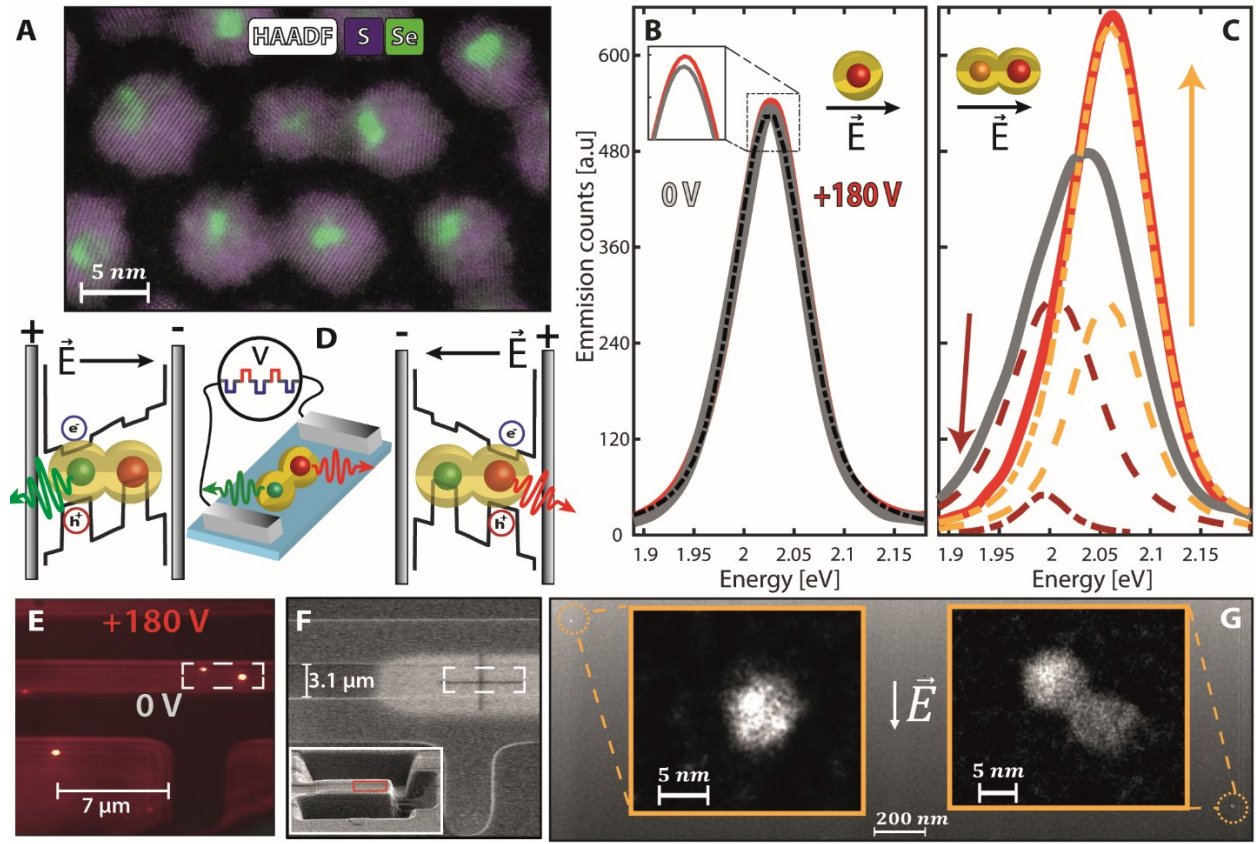

**Figure 1. Electric field modulation of fluorescence in dimer-CQDM versus monomer-QD.**

**a,** STEM-HAADF image of the dimer sample with overlay of the elemental mapping by Energy-dispersive X-ray spectroscopy (EDS) of Selenium (green) and Sulfur (purple), showing the core/shell structure. (full elemental analysis in Supplementary Fig S1). Dimer-CQDM to mono-QD percentage in the sample is ~85%/15%. **b-c,** Photoluminescence (PL) emission spectra for each particle, with (red) and without (gray) an EF (2 minutes integration time). The mono-QD PL is fitted by a single Voigt function with peak energy, $PE = 2.025 \pm 0.002\ eV$ and $FWHM = 85 \pm 3\ meV$, to which the EF induces minor intensity modulation of the PL (inset). The CQDM emission can only be fit by 2 separate Voigt peaks (without field: dashed lines, red peak: $PE_R(0V) = 2.005 \pm 0.006\ eV$; $FWHM_R(0V) = 106 \pm 9\ meV$, yellow peak: $PE_Y(0V) = 2.056 \pm 5\ eV$; $FWHM_Y(0V) = 88 \pm 5\ meV$). With field: dash-dot lines, red peak: $PE_R(180V) = 1.997 \pm 0.002\ eV$; $FWHM_R(180V) = 60 \pm 20\ meV$, yellow peak: $PE_Y(180V) = 2.060 \pm 0.003\ eV$; $FWHM_Y(180V) = 95 \pm 3\ meV$). **d,** Schematic of the CQDM color switching device and mechanism. The CQDM is excited and generates single photons from both color centers of the system (center). Upon external electric field application, excitons are pushed to the QD opposing the field direction due to the electron toggling. **e,** Optical widefield PL image showing the excited single particles from B-C (white dashed rectangle) between two electrodes. The outer electrodes are also designed with ruler marks to later correlate optical with electron microscopy structural measurements. **f,** SEM image of the same area as in E. The desired area is coated with a protective layer and then cut out of the device (inset), to produce a lamella that is placed on an electron microscope grid. **g,** STEM-HAADF image of the area of interest shown in E-F, (dashed white rectangle). Two NCs are seen in close up- correlated with the locations in the PL image in (e), showing a single QD (left) and a fused CQDM (right). The direction of the E-field is also indicated.

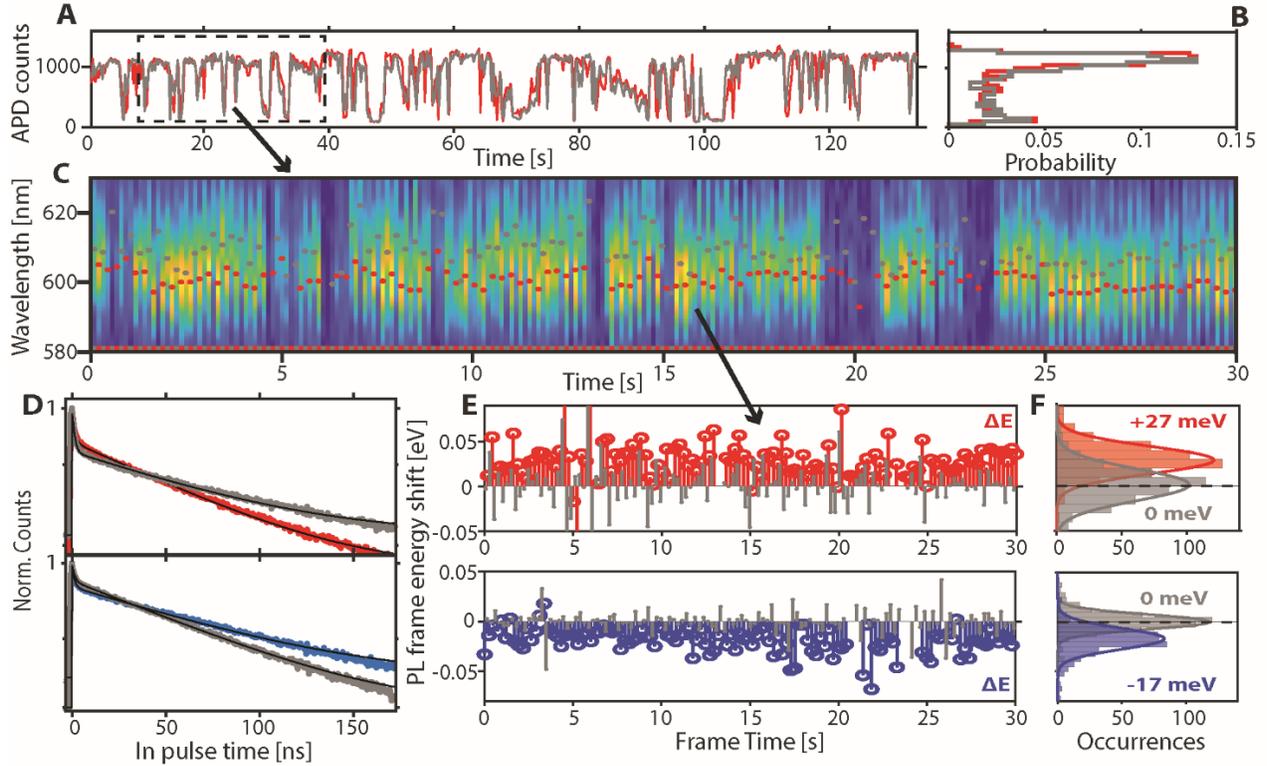

**Figure 2. Electric field induced color switching at 10 Hz modulation rate.**

**a,** Time trace of PL intensity for the CQDM shown in Fig. 1c, and **b,** it's probability histogram for time frames with (red) and without (gray) applied electric-field. **c,** A 30 second strip of successive spectra (100ms bins). The gaussian fitted PL peak position at each frame with/without EF modulation is indicated (red/gray dots). **d,** Normalized lifetime decay traces with/without (color/gray) EF for the +180V to 0V (top frame, red/gray) measurement shown in Fig 1c, and a successive measurement of -180V to 0V (bottom frame, blue/gray) which also shows a shorter lifetime without EF, probably due to surface charging of the particle, which occurs throughout the measurements. Black lines are fits to a double exponential decay. We note that for this particle each ~2-minute measurement consisted of a single EF voltage modulation, and the CQDM exhibits change in the 0V PL spectra and intensity between measurements at different applied EF values (details in Supplementary Fig. 9). **e,** The energy shift between subsequent frames $\Delta E^{+180V-0V}$ (red) and $\Delta E^{0V-0V(next)}$ (gray) from the first measurement (top box), and between subsequent frames $\Delta E^{-180V-0V}$ (blue) and $\Delta E^{0V-0V(next)}$ (gray) from the measurement with the opposite EF polarity. **f,** Histograms of the data in E fitted by a normal distribution, the EF modulated frame energy shifts (red, blue) exceeds the maximal $\Delta E^{0V-0V(next)}$ frame-shifts originating from the spectral diffusion phenomenon.

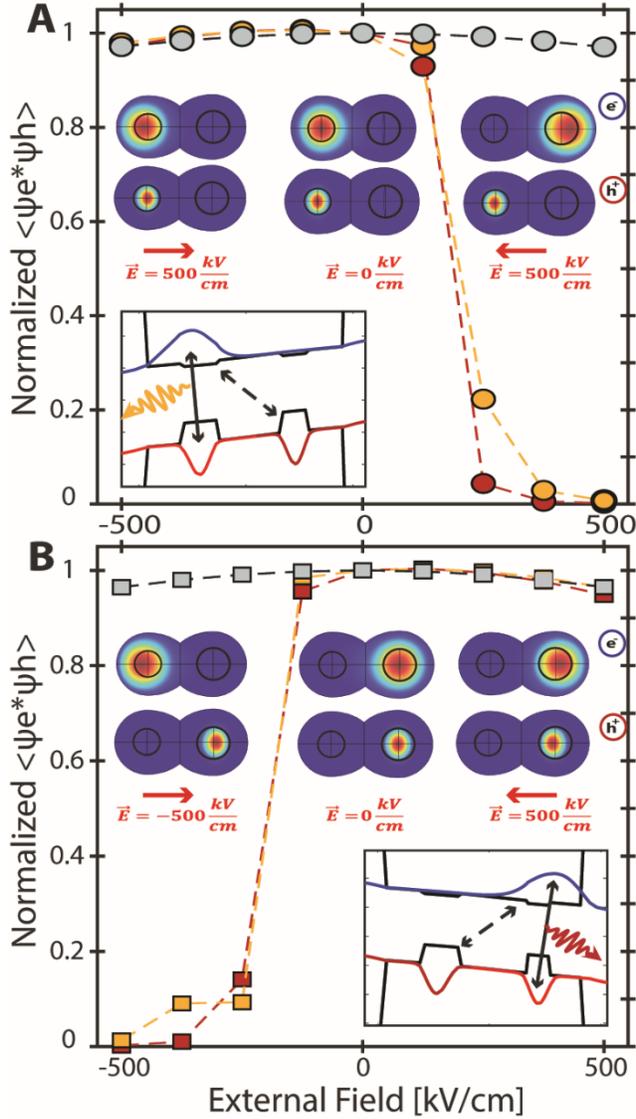

**Figure 3. Effective mass calculations of electron and hole states under applied electric field**

**a,b:** Calculated electron-hole wavefunction overlap under applied external EFs at different orientations between ±500 kV/cm, normalized to the 0 kV/cm value (0.67/0.72 for the yellow/red emission center, respectively), in a CQDM with yellow/red localized hole, a/b respectively. (0° (red), 45° (yellow), and 90° (gray) relative EF orientation to the CQDM fusion axis are shown). Inset images represent the electron and hole wavefunction densities calculated for different electric fields. Square insets present the band structure with the electron and hole ground states at ±500 kV/cm, showing direct - radiative (full line) and indirect - non-radiative (dashed) relaxation pathways.

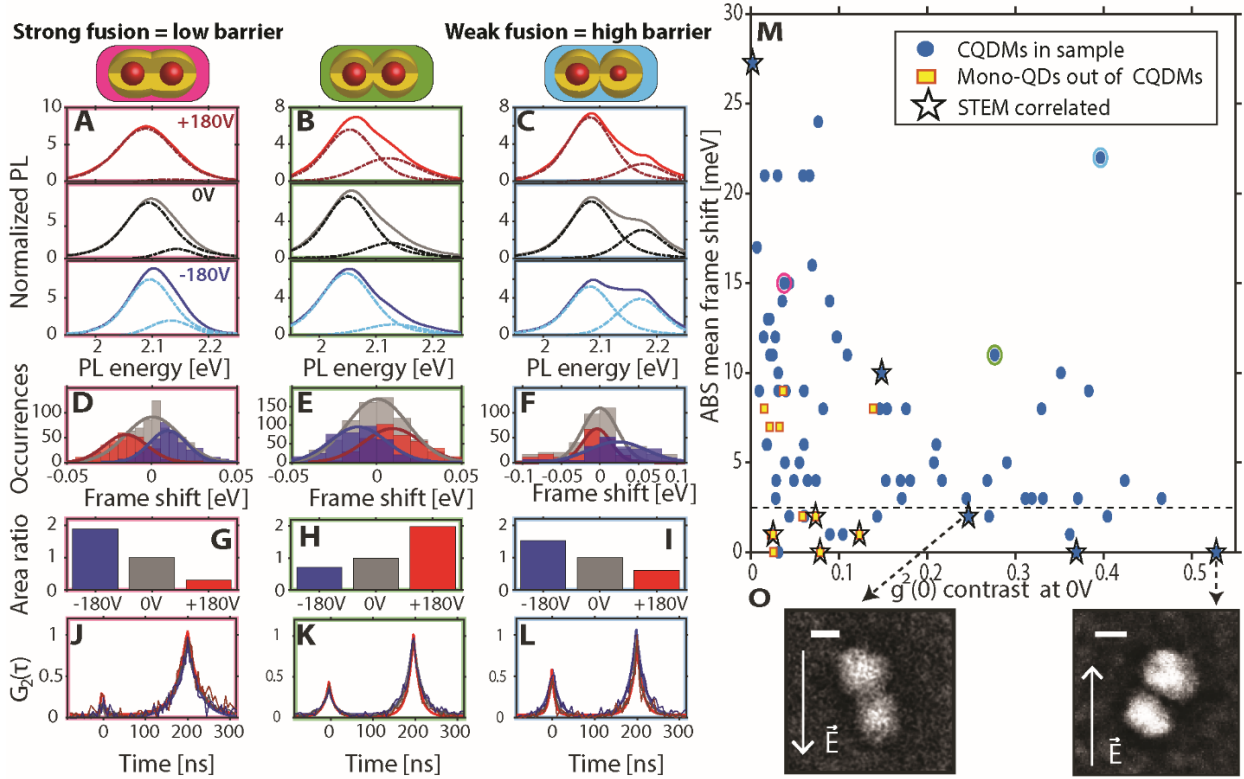

**Fig. 4 correlation between g²(0) and color switching.**

**a-c,** Are the accumulated PL spectra at +180V (red), 0V (gray) -180V (blue) of selected CQDMs (colored circles in M), with 2 peak Voigt fitting (dashed). **d-f,** Histograms of the energy shift between subsequent frames $\Delta E^{+180V-0V}$ (red), $\Delta E^{0V-0V(next)}$ (gray) and $\Delta E^{-180V-0V}$ (blue), with normal distribution fitting for each particle. **g-i,** A comparison of the area ratio of the "high" to "low" energy-peaks from the fitted spectra in A-C, normalized to the ratio from the 0V spectra (gray). **j-l,** Normalized G²(τ) second order photon correlation traces for the EF modulated measurements, from which g²(0) is calculated for each applied voltage. **m,** Scatter plot showing the mean energy frame-shift upon EF modulation (taken as the fitted maxima in D-F, in absolute value) and the 0V g²(0) values of 89 single particles consisting of 77 CQDMs (circles) and 12 mono-QDs (squares). Stars indicate particles later correlated with STEM measurement. Colored circles indicate the CQDMs shown in full analysis in A-L (magenta- "strongly fused", green- "medium fusion regime", and cyan- "weakly fused hetero- dimer"). **o,** STEM images of high g²(0) CQDMs which exhibit low EF effect due to the weak coupling between emission centers, indicated by the high g²(0).

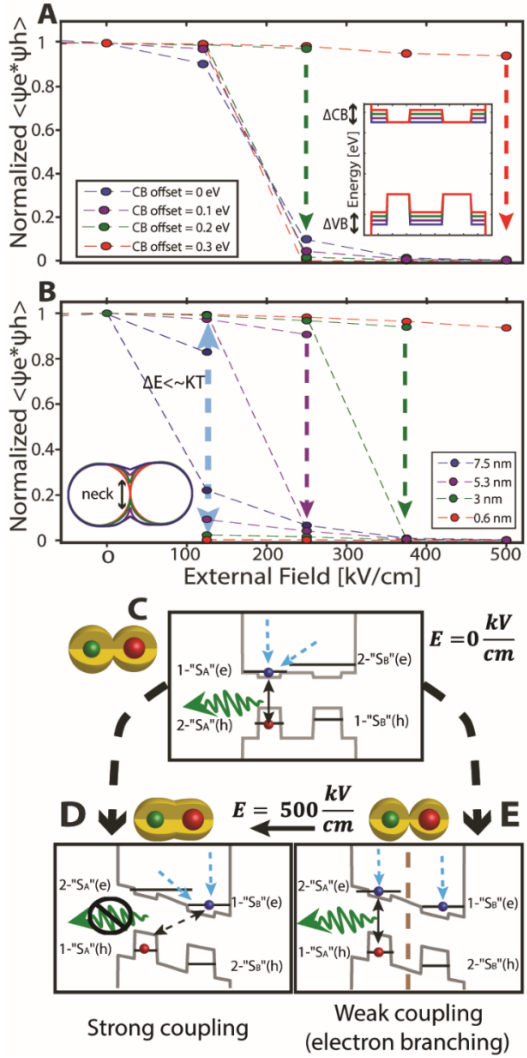

**Fig. 5: Calculations of color switching for different band offsets and different neck widths (fusion).**

**a,** The Calculated electron-hole wavefunction overlap in a CQDM geometry under applied external EF normalized to the 0 kV/cm value, for conduction band offsets ranging between 0 to 0.3 eV. **b,** Same calculation with a fixed conduction band offset of 0.1 eV and changing the CQDM fusion neck-width from 7.5 nm (dark blue) to 0.6 nm (red). For both A-B the hole is calculated in the 1-"S$_A$" state in the smaller ("green") core. Colored arrows (color code as in legends) in A-B indicate the probable occurrence of a transition between 2-"S$_A$" to 1-"S$_B$" states for the electron, presumed at an inter-core electron wavefunction delocalization higher than 1%. Light-blue double arrow at 125 kV/cm in B indicates when the energy separation between the 2 electron states is below KT at room temperature, resulting in thermal population mixing (occurring for all but the 0.6 nm neck width). **c-e,** Schematic band diagram showing the probable electron relaxation pathways for the electron when the hole is localized in the smaller core without applied EF (C), with 500 kV/cm EF and at the strong (D) or weak (E) coupling limits between the QDs.

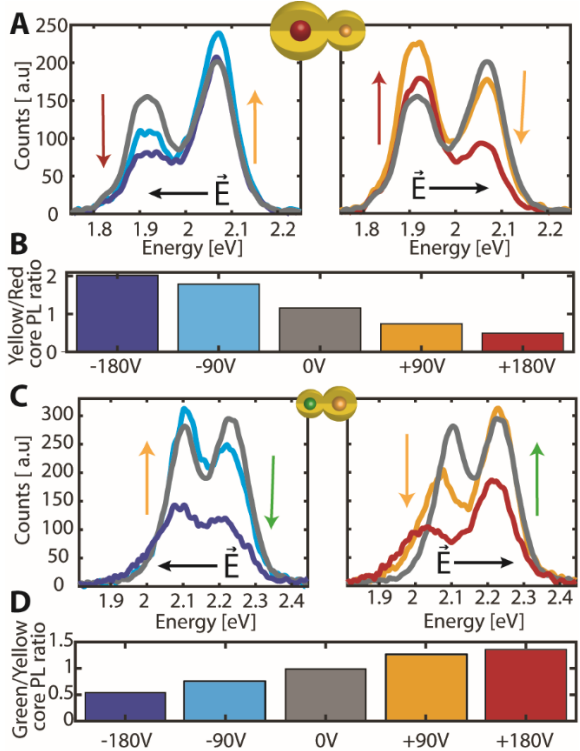

**Fig. 6: Broadband color switching in hetero-sized CQDMs.**

**a,** Accumulated PL spectra of a CQDM with distinct emission centers at 1.917 eV ("red") and 2.065 eV ("yellow"), that shows broadband color switching. An electric field in the direction of the "red" center (-180V, -90V, left panel) switches to preferred emission from the "yellow" center, while an opposite polarity EF (+180V, +90V, right panel) switches the emission to the "red" center. **b,** the PL population ratio between the two peaks ($\frac{\int PL_{yellow}}{\int PL_{red}}$) is calculated from the two-Voigt fit of each spectrum (supplementary Fig S26), showing the emission probability of the two peaks is nearly equal without applied EF (gray), the "yellow" center emits twice (half) as much as the red with -180V (+180V) applied EF. **c,** Accumulated PL spectra of a CQDM with distinct emission centers at 2.100 eV ("yellow") and 2.235 eV ("green"), that also shows broadband color switching without intensity loss at ±90V applied EF. At ±180V the PL color switching is accompanied by intensity dimming along with red shifting and line broadening of both peaks, which indicates that the applied EF induces also meaningful electron-hole charge separation. **d,** the PL population ratio between the two peaks ($\frac{\int PL_{green}}{\int PL_{yellow}}$) shows the same trend of PL population switching between emission centers induced by the magnitude and polarity of the applied EF.

## **Methods:**

Synthesis of CdSe/CdS colloidal CQDM:

The formation of fused CdSe/CdSe CQDMs includes several steps and is described in our prior works[21,31]. Starting by (1) synthesis of CdSe/CdS core/shell QDs with the desired size and (2) preparing $SiO_2$ nanoparticles (NP) and using them as a template. Next, (3) binding the QDs to the silica surface and masking the NPs with an additional layer of $SiO_2$ and, (4) Binding a molecular linker followed by adding more QDs, creating linked CdSe/CdS dimers on the template. Finally, (6) etching the silica using HF and fusing the dimer in a Schlenk-line connected flask for 12 hours at 240C°. For achieving high yield of CQDMs, post synthesis size selective precipitation is done as in our previous works, sorting the CQDMs out of the mono-QDs. The separation can be repeated until the desired purity is achieved. Based on this procedure the different CQDMs in this work were made.

The formation of "red"/"yellow" CQDMs was achieved using two-different sizes of QDs 2.2/5.3 nm (red QDs) and 1.2/4.8 nm (yellow), respectively. After binding the "red" QD to the silica template the "yellow" QD was attached using the molecular linker. The "yellow"/"green" CQDMs were made using nominally 1.4/2.1 nm QDs with broad size distribution. Based on their heterogeneity, a hetero-CQDM with two emission centers were formed in a statistical manner.

Fabrication of Horizontal electrode devices

$24 \times 60$ mm$^2$ Borofloat 33 glass cover slides of 175 µm thickness were cleaned in a Piranha solution for 10 minutes. A 200 nm Aluminum (Al) layer was evaporated (Angstrom Quantum Series Evaporator) on the cleaned substrates. Etching of the Al layer into the device design was done by a photolithography process using a laser writer system (Microtec LW405B) or a mask aligner (MA6, SUSS Microtec) and dry etching the Aluminum layer (RIE System, Plasma Lab 100, Oxford Plasma Instruments) to produce electrode arrays of ~3 µm width and 3 µm spacing. The outer electrodes in each array are designed with external ruler markers and numbering, allowing to spatially resolve areas of interest with measured particles, seen in the wide-field PL image, and find them later in a Focused Ion beam scanning electron microscope (FIB-SEM).

Single particle optical measurements

Single particle PL measurements were performed with an inverted microscope (Nikon Eclipse-Ti) in epi-luminescence configuration. The PL signal from single separate particles was collected through an oil immersion objective (×100; 1.4 NA Nikon Achromat), which was also used for focusing the excitation light: a 370 nm fiber coupled widefield LED (Prizmatix) for imaging, or a 450nm pulsed picosecond laser (EPL450, Edinburgh instruments) at 5 MHz repetition rate, for single particle measurements. A dilute solution of CQDMs in 2wt% poly(methyl-methacrylate) (PMMA) in toluene, were spin coated on the house fabricated electrode devices at 4000 RPM, creating an 80-100 nm thick film with single separated CQDMs, evenly spread out between the electrodes. The broad-sized small CQDM sample shown in supplementary figure S25 A,C,E, were dispersed in 1.3% Polyvinylpyrrolidone (PVP) in chloroform, and spin coated on 2.5 μm spaced electrodes. PVP shows a much higher dielectric constant[56] (~7.7) than PMMA, leading to an enhanced EF effect on the electron as seem in figure 6 and supplementary figures S28-31.

The electric field (EF) between the electrodes is generated by a waveform generator (Agilent 33220A) producing a periodic DC electric potential step signal which is amplified ×50 (9100A Tabor electronics) and connected to the electrodes on the substrate. Pre-measurement, a widefield image of a 40×40 μm area is taken with an Electron-Multiplied-Charge-Coupled-Camera (EMCCD, IXon 3, Andor instruments, ProEM1024, Princeton instruments for samples in figure 6 and supplementary figures S25-S31) to spatially resolve the single particles on the electrodes. A single CQDM is excited with the EPL450 laser with circularly polarized light at an average exciton per pulse value of $< N > \sim 0.2$. The emission light is passed through a dichroic mirror (DMLP490R, Thorlabs) and an additional long pass filter (550LP, Thorlabs) before passing through a motorized λ/2 waveplate (KPRM1E/M, with AHWP05M-600 waveplate, Thorlabs). The emission is then split through a 50:50 beam splitter (Thorlabs), with one part passing through a polarization beam displacer (BD40, Thorlabs), into a spectrograph (SP2150, Princeton instruments) and imaged on the EMCCD. The second part of the emission is focused onto two Avalanche Photodiodes (APD, Excelitas SPCM-AQRH-14) in a Hanbury-Brown-Twiss (HBT) geometry. The measurement start is triggered by a positive voltage pulse, and the spectra is measured in a kinetic series of 1000-1600 frames (100 ms accumulation each), with different applied EF modulated periodically between adjunct frames.

The electric field modulation is done at either a 2-step sequence (+V,0V) for different applied voltage, or a 4-step sequence (+180V, 0V, -180V, 0V) to detect one sided or dual sided color switching of the emission. The spectra are analyzed both at single frame level and total measurement accumulation, smoothed by a gaussian filter to average out the readout and Shot noise of the camera. The $\lambda/2$ waveplate is rotated at 9° steps between 0°-135° relative to the electrodes long dimension, allowing us to calculate the emission's linear polarization angle and degree from the intensity of the horizontal (H) and vertical (V) polarized spectra on the EMCCD (additional explanation in supplementary Fig. S3)[34]:

$$P(\theta) = \frac{I_H(\theta) - I_V(\theta)}{I_H(\theta) + I_V(\theta)} = P_{degree} \cdot \cos(4\theta + \phi)$$

Where $I_H$ and $I_V$ represent the displaced horizontal and vertical polarized emission intensity, $\theta$ is the rotation angle of the $\lambda/2$ waveplate (with 0° being parallel to the electrodes), $P_{degree}$ is the degree of circular-linear (between 0-1, respectively) emission polarization and $\phi$ being the emission's linear polarization angle with respect to the electrodes (between 0°-180°).

Photon statistics of the signal from the Avalanche Photodiode detectors is performed using a multichannel Time Tagger 20 (Swabian Instruments), to which the timing of the voltage modulation, camera frames, and $\lambda/2$ waveplate rotation times are also registered simultaneously, allowing to separately analyze the behavior at each voltage step.

The PL spectra are accumulated from the measurement separately for each voltage step, and fit to a single or double Voigt line function for a single/dual color emitting CQDM.

The emission spectrum is also analyzed at the single frame level by a gaussian profile fit finding the peak emission energy $E_{frame}^{V_i}(n)$ ,$V_i$ is the voltage step and n the cycle of EF modulation. The energy shift induced by the external voltage step between frames is found by

$$\Delta E^{V_i}(n) = E_{frame}^{V_i}(n) - E_{frame}^{0V}(n)$$

and is compared to the energy shifts between frames without EF modulation:

$$\Delta E^{0V}(n) = E_{frame}^{0V}(n) - E_{frame}^{0V}(n+1)$$

which are induced by spectral diffusion events. The "mean frame energy shift" $< E_{frame}^{V_i} >$ referred in the main text is the mean of the normally distributed fitting of $\{\Delta E^{V_i}\}$ (shown in Fig. 2f and Fig. 4d-f), and is compared to $< E_{frame}^{0V} >$ which shows a mean frame shift of 0 meV due to random charging events within the NC [39–41].

Second order correlation function of photons in the HBT setup (which receive 25% of the photons that went through two 50:50 beam splitters) is found separately for measurement times with different applied EF, with the $G^2(\tau)$ contrast or "photon purity"[45] calculated by the area of the $G^2(0)/G^2(200ns)$ peaks (fitted by $G^2(t) = A_0 e^{-k_1|t|} + A_{200} e^{-k_2|t-200|}$, where A are the amplitudes and k are the rate constants of the multi (t=0) and single (t=200) photon emissions.

The fluorescence lifetime is fit to a multi-exponential model (up to 3 exponents) to account for short, intermediate, and long timed recombination pathways. An assumption for this model and the $G^2(\tau)$ contrast is that the neutral mono-exciton quantum yield is 100%, which is a good assumption for the bright state of this type of core-shell single-NCs[40,57].

The time dependent spectrum, polarization degree, photon time traces, fluorescence lifetime, and second order photon correlation were extracted from the time-tagged, time-resolved, voltage-resolved data by using a home written MATLAB code.

Spatially resolved single particle electron microscopy (see movie S1)

Selected single particles are spatially resolved on the substrate from the widefield optical pictures. The electrode cover-slide device is taken to a scanning electron microscope (SEM) equipped with a Dual Beam Focused Ion Beam (DB-FIB, Helios Nanolab 460F1Lite Dual Beam Focused Ion Beam/Scanning Electron Microscope ThermoFisher, former FEI) that allows nanoscale ion milling of the device as well as amorphous Carbon and Platina deposition. The selected area on the device (typically a $3.5 \times 10 \ \mu m^2$ area between electrodes) is covered first by a 250nm amorphous carbon layer, then a 1 $\mu m$ Pt layer, then the area around it is milled by $Ga^+$ ions.

The piece is then tilted on the flip stage and lift out with by the EasyLift Nanomanipulator, and the carbon and Pt layers are milled off to create a thin TEM lamella (typically 50 nm thick glass base) which is put on a FIB lift out TEM grid (Ted Pella). This grid is then scanned using an Aberration Probe-Corrected Scanning Transmission Electron Microscope (Themis Z G3, ThermoFisher, former FEI), using a High-Angle-Annular Dark Field (HAADF) detector at low electron current settings. Single nanocrystals are seen in bright contrast by HAADF imaging on the lamella background from 1nm pixel size, and close up images of <1Å pixel size show size and shape resolved single CQDMs, which can be correlated to the optical measurements, where single particles are identified by widefield PL imaging.

Numerical calculations of the charge carrier states

All numerical simulations were done using COMSOL Multiphysics version 5.6. The model of a 3D CQDM is built using the HAADF image of the nanoparticle to determine the 2D ellipse shape of each QD, and the fusion neck width between them. The CdSe core size in each QD is taken as a sphere size which gives exciton energies fitting the peaks of the accumulated emission spectra. The wavefunctions and energies of each charge carrier state (hole or electron) are calculated by solving the self-consistent Schrödinger-Poisson equation as shown in our previous works[21,28,35] (self-consistent iteration sequence further explained in Supplementary text S1) taking into account the coulomb interaction between the two charges and the self-potential, and adding an external electric field induced potential by solving the Laplace equation on a 100 nm$^2$ domain:

$$\nabla(\varepsilon(r)\nabla)V_E = 0 \big|_{V_E(z=0)=0 \text{ V}}^{V_E(z=100 \text{ nm})=V_{ext}}$$

Where $\varepsilon(r)$ is the relative dielectric constant which induces dielectric screening effects between the NC and the surroundings, and $V_E$ is the potential induced by the EF with Dirichlet boundary conditions that simulate the EF magnitude. The electron and hole wavefunctions and energies of the ground state in each core- "$S_{A/B}$" are calculated as described in supplementary text S1. Iterations are done until the energy difference between subsequent iterations converge, usually after 3-4 cycles of the calculation. In a CQDM, we get two ground states for the electron and two for the hole, with the wave function hybridization and energy depending on the system dimensions (core size, shell thickness, neck width) and coulomb binding energy between charges. Additionally, we calculate the overlap integral for an electron-hole pair $< \Psi_e^* \cdot \Psi_h >^{EF}$ under the potential of an applied electric field, and normalize it to the value of $< \Psi_e^* \cdot \Psi_h >^{EF=0}$ as a measure of the relative emission modulation by an applied electric field, as well as the normalized wavefunction probability density in each QD. The computational space extends more than 2 times the CQDMs size in each direction with the Dirichlet boundary-condition setting the wavefunction to zero at the edges. We use von Neumann boundary-condition at the inner (between core-shell) and outer boundaries of the QD in order to impose the Ben-Daniel-Duke condition. The applied electric field between the two electrodes is also calculated by solving the Laplace equation on the specific area geometry, considering the dielectric polarization effects of the surroundings (glass, substrate, PMMA, air). The result is an almost uniform field (linearly changing voltage potential) parallel to the substrate plane between the electrodes. For the CQDM shown in figure 1 we calculated a field of 460 kV/cm by the 180V applied potential between the electrodes (supplementary Fig. 10).

We consider the two ground excited states (in either of the two emission centers) for the electron and hole as possible recombination pathways of excitons. The higher energy state of each charge carrier is presumed to be metastable if the wavefunction density in the opposite center is less than 1%, otherwise it is presumed that the higher state will relax to the lowest possible energetic state (sub-Pico to Pico-second process between mono-QD 1P-1S states with high wavefunction overlap[58–61], and sub nano-second process for excitonic energy transfer[62] with low coupling between emission centers), faster than the radiative recombination timescales (nano-seconds). When an energy difference between two states is less than KT in room temperature, the states are presumed to be simultaneously populated, which allows possible radiative recombination from both of them.

**Supplementary Information**

for

**Electric field induced color switching**

**in colloidal quantum dot molecules at room temperature.**

Yonatan Ossia[1,2], Adar Levi[1,2], Yossef E. Panfil[1,2], Somnath Koley[1,2], Einav Scharf[1,2], Nadav Chefetz[1],

Sergei Remennik[2], Atzmon Vakahi[2] & Uri Banin[1,2]*

[1]Institute of Chemistry, [2]The Center for Nanoscience and Nanotechnology, The Hebrew University of Jerusalem,

Jerusalem 91904, Israel

*Corresponding author: Prof. Uri Banin, uri.banin@mail.huji.ac.il

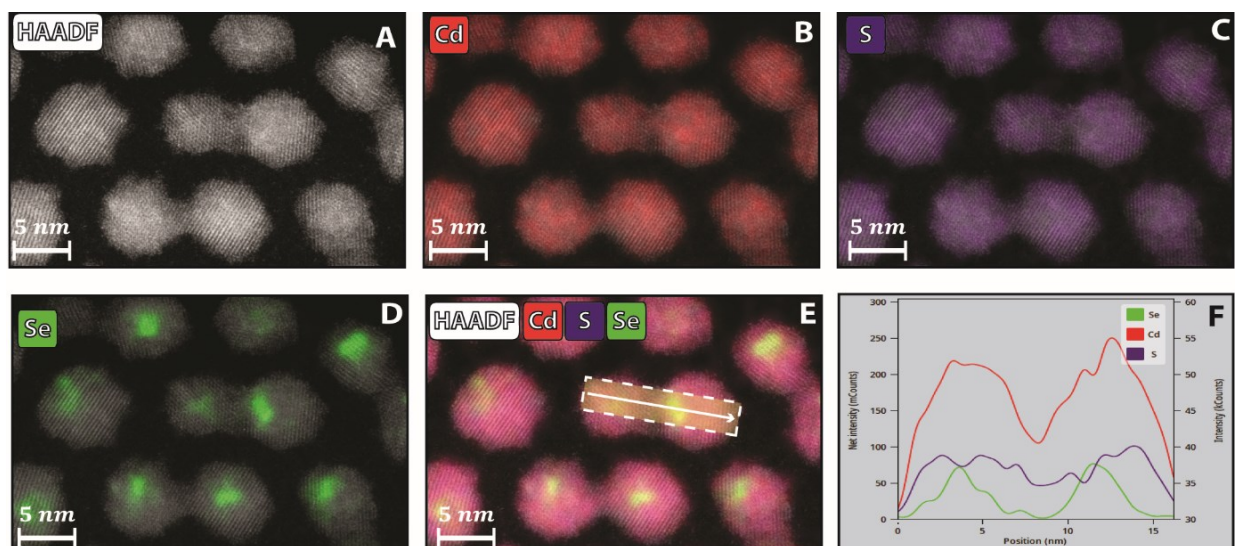

**Figure S1. Structural characterization and analysis of studied Colloidal quantum dot molecules (CQDMs).** (A) High-Angle-Annular-Dark-Field Scanning-Transmission-Electron-Microscope (HAADF-STEM) image of coupled/single QDs from the sample used in all experiments (approximate 85% CQDMs), showing a distribution in QD size and inter-particle fusion degree (neck width). STEM-EDS analysis of Cadmium (B), Sulfur (C), Selenium (D), and a combined Energy-Dispersive-Spectroscopy (EDS) elemental analysis (E) line area scan data (F) show distinct CdSe cores coated by a CdS shell.

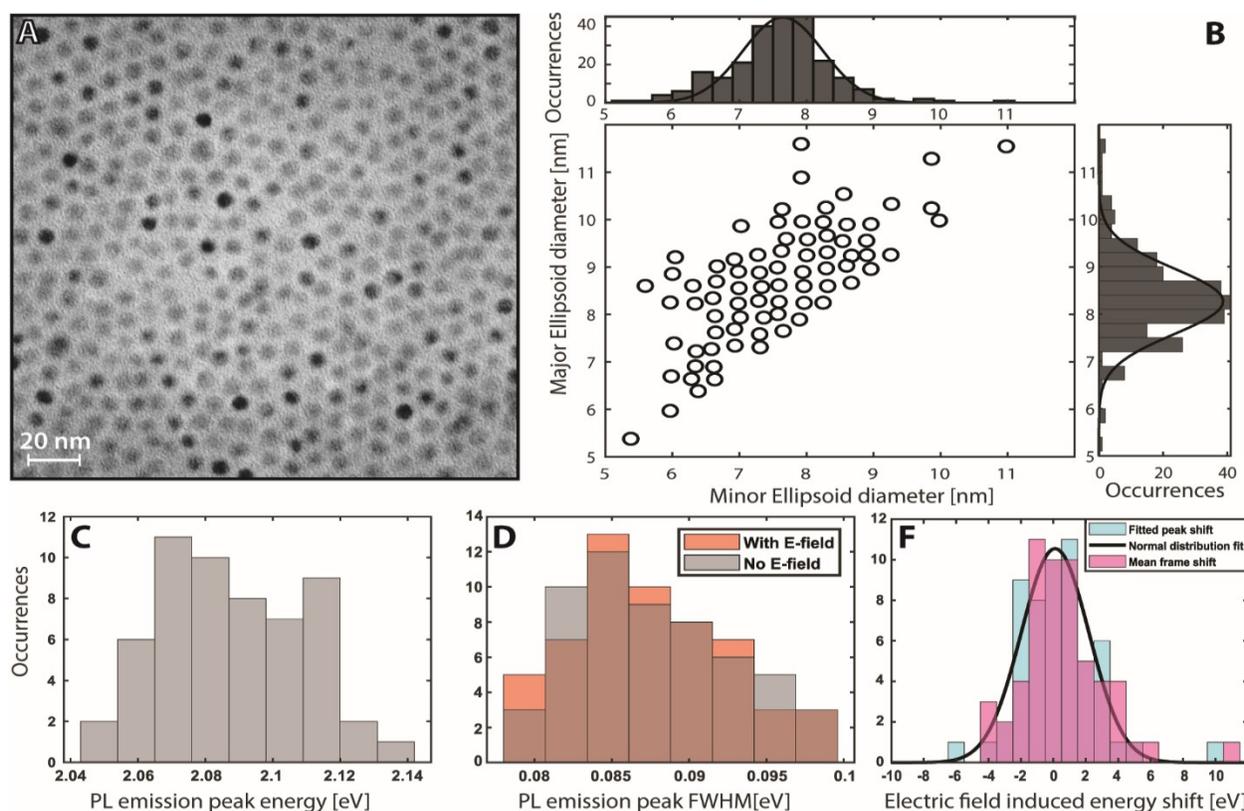

**Figure S2. Statistical data of mono-CdSe/CdS QDs measured with/without applied electric field (EF). (A)** Transmission electron microscope (TEM) image of mono-QDs on an ultrathin carbon TEM grid. **(B)** Size statistics on 240 QDs show a slightly elliptical final structure with an average major/minor axis of 8.3±1.6 / 7.7±1.4 nm, respectively. The axis error distribution is taken as the full width at half maxima (FWHM) of the normal distribution fit to the data. **(C)** The emission photoluminescence peak energy from 47 single QDs measured in the same setup as in Figure. S3 shows that the QD size distribution results in an emission energy distribution of up to 100 meV. **(D)** The emission spectra FWHM for mono-QDs also ranging between 75-100 meV, with negligible changes with/without applied EF. **(F)** The emission energy shift of these single QDs, calculated by the accumulated spectra peak shift or by the mean frame shift analysis as explained in "methods", is normally distributed around 0 meV. This indicates the main parameter affecting the EF response is the random orientation of the QD with respect to the EF direction[1–3]. The standard deviation is ±2.5 meV, for both the accumulated emission peak shift and the mean frame analysis shift with/without applied field. The maximal observed emission shift was 10 meV, seen for the QD in figure S8 for which the C-axis is aligned in parallel to the electric field, as presumed from the PL polarization measurement[4,5].

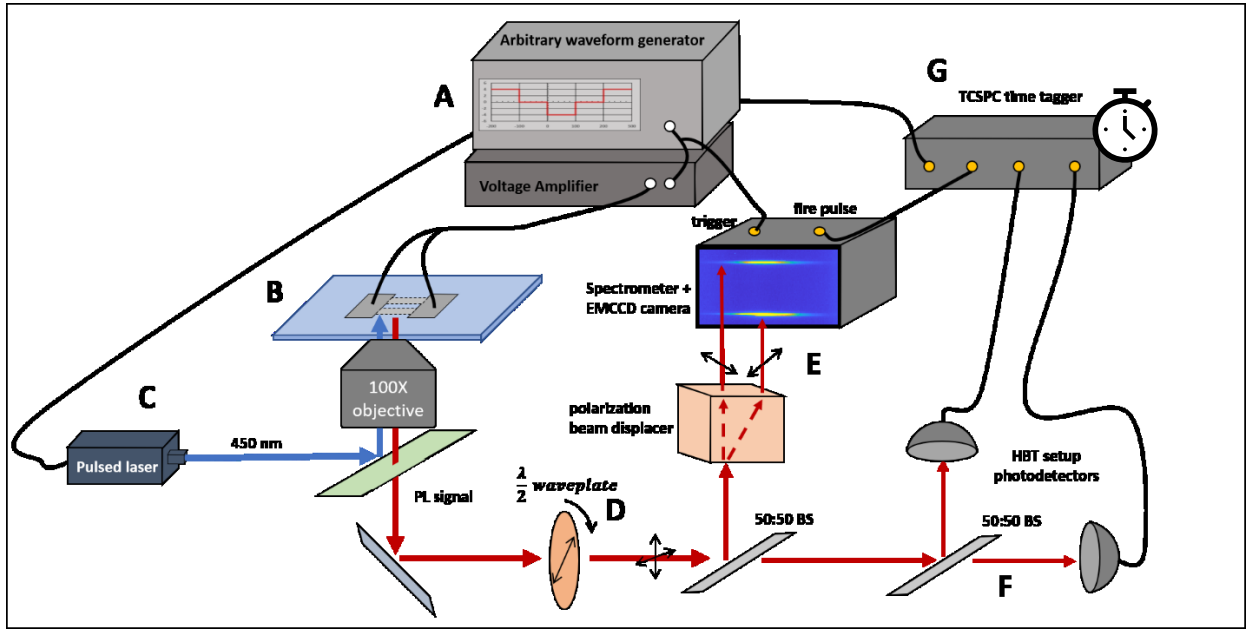

**Figure S3. Schematics of the electro-optical experiment setup**. (**A**) An arbitrary waveform generator (Agilent 33220A) produces a periodic DC electric potential step signal which is amplified ×50 (9100A Tabor electronics) and connected to the electrodes on the substrate mounted on (**B**) an inverted microscope setup (Nikon- ECLIPSE Ti) with an oil emersion ×100 objective (Nikon CFI Plan Apochromat Lambda D). Single particles between the electrodes are excited with a (**C**) 450 nm picosecond laser (EPL450, Edinburgh instruments) at 5 MHz repetition rate. The PL signal is passed through a (**D**) $\lambda/2$ waveplate that rotates throughout the measurement at 9° every 7 seconds, and is aligned respecting linear polarized light parallel to the electrodes. The PL signal is passed through a 50:50 beam splitter (BS) (Thorlabs) with the first portion going through a (**E**) polarizing beam displacer to a SPI150 spectrometer (Princeton instruments) connected to an electron-multiplying-charge-coupled-device (EMCCD, IXon 3, Andor), taking measurements at 10 Hz rate, giving the PL spectra for each frame, and PL polarization degree and angle. The second portion of the PL is sent to a (**F**) Hanbury Brown and Twiss (HBT) setup with 2 single photon counting modules (Excelitas SPCM-AQRH-14), in which we measure the PL lifetime and second order coincidence function $G^2(\tau)$. All measurement components are synchronized by a time-correlated-single-photon-counter (Time tagger 20, Swabian instruments) (**G**) to which the EMCCD frame times and DC voltage modulation are also recorded.

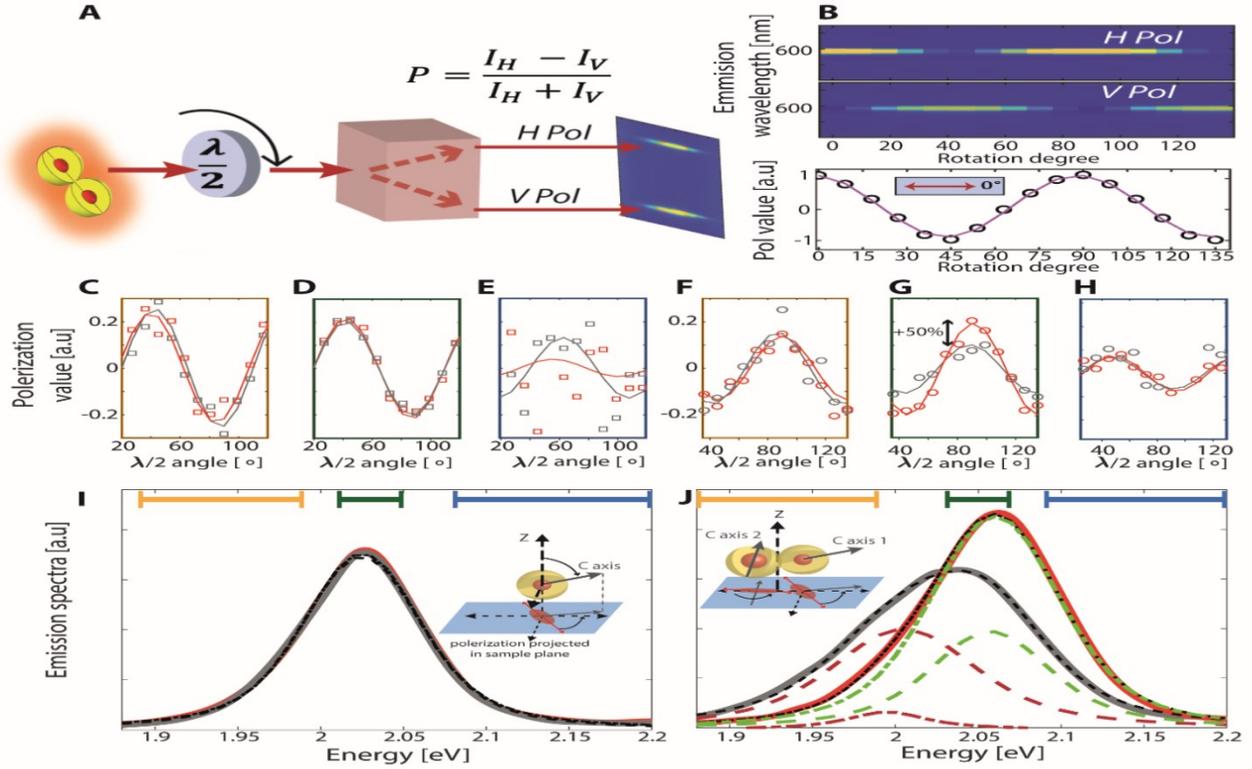

**Figure S4. Emission polarization and spectrum correlation in mono-QDs and CQDMs**. (**A**) Demonstration of emission polarization value and angle measurement. The emission spectra on the EMCCD is separated into its horizontal (H) and Vertical (V) polarized components with respect to the electrodes. the polarization value is calculated by the equation: $P(\lambda) = \frac{I(\lambda)_H - I(\lambda)_V}{I(\lambda)_H + I(\lambda)_V}$ where I(λ) is the integrated intensity of the emission on the camera for a selected wavelength range. (**B**) The calibration measurements of 590nm linearly polarized light passed parallel to the electrodes. The V&H emission (upper color maps) are taken at 9° rotation steps of a λ/2 waveplate. The polarization values as a function of rotation degree (bottom plot) is fitted by the function: $P = P_0 * \cos(4\theta + \phi)$ where $P_0$ is the degree of linear polarization (equals 1±0.05 in the calibration fit for polarized light), θ is the rotation degree, and ϕ is the angle between the polarized light and the electrodes (equals 0°±5° in the calibration light fit). (**C-H**) Plots of emission polarization values along with fitting for the single mono-QD (**C-E**) and CQDM (**F-H**) at three different spectral ranges with (red) and without (gray) Electric-field (EF). The polarization values and angles for three ranges are extracted from the low energy (C, F, yellow), the center (D, G, green), and the high energy (E, H, blue), part of the total accumulated emission spectra. The Mono-QD exhibits a constant ϕ angle of 84°±5° for the entire spectrum (the blue part is fitted poorly, as these photons are emitted from charged states and fine structure excited states, which have a different polarization from the ground state[5]) . The dimer shows ϕ angle of 177°±5° for the red and center parts, and 94°±5° for the blue part of the spectrum. The center part's $P_0$ value changes from 0.1 to 0.2 with EF (**G**), indicating that the source of emission shifting is indeed shifts in the emission probability of two distinguishable cores. (**I-J**) Total accumulated spectra showing the three parts taken in polarization value calculation for the mono-QD (I) and CQDM (J). The CQDM spectrum is deconvoluted to exhibit the contribution of the two cores

in the accumulated emission spectra (dashed lines). Inset shows how the a-b plane emission polarization is projected to the sample plane according to the C-axis orientation.

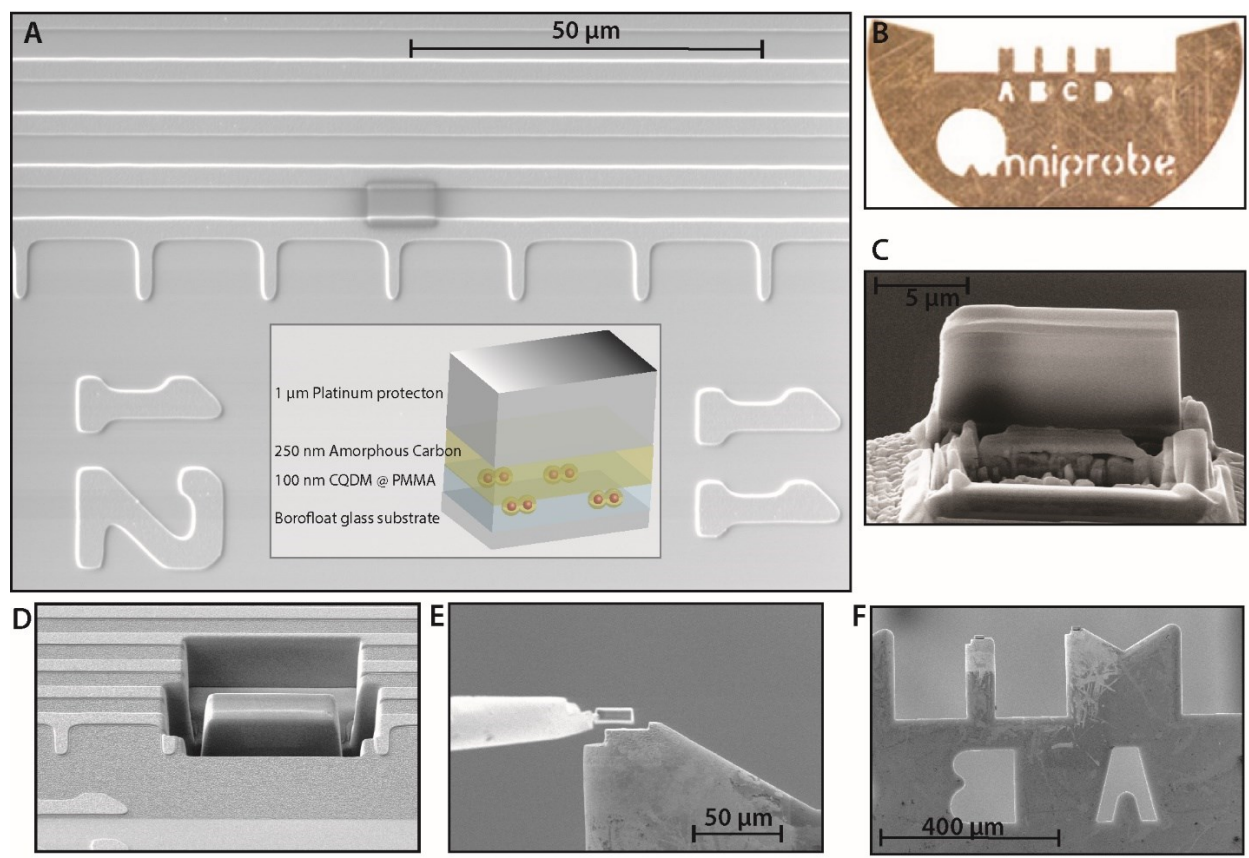

Figure S5. Post measurement electrodes SEM image and electron microscope grid preparation. (**A**) SEM image of interlocked aluminum electrodes with ruler marks and numbering on the outer electrodes. Between the bottom two electrodes is a region of interest (rectangle) covered with amorphous carbon, as part of the protection layers (scheme in inset) deposited by the Dual-Beam-Focused-Ion-Beam (DB-FIB) before cutting out a lamella to be put on a FIB-lift out TEM grid (**B**). (**C**) The lamella on the TEM grid still showing the edges of the interlocked electrodes. (**D**) The region of interest is cut out from the sides with the Ion beam, then connected to a Nanomanipulator and moved to the FIB-lift out grid (**E**) and connected (**F**) for further imaging by High resolution scanning transmission electron microscopy.

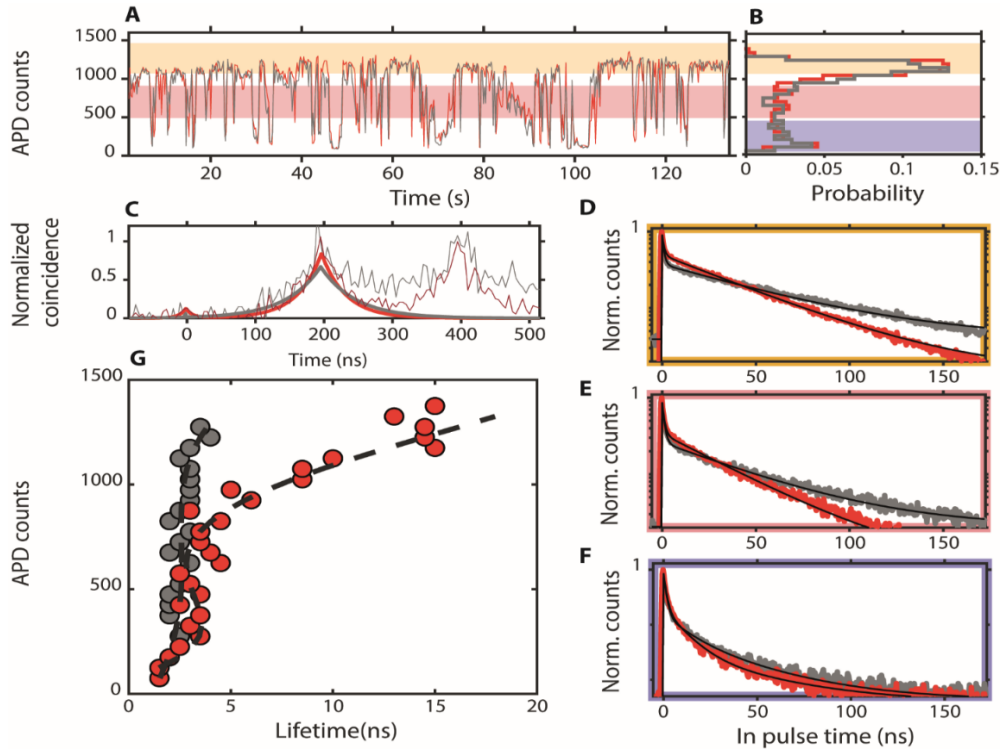

**Figure S6. Photon counting statistics for the CQDM in main article figure 1. (A)** emission intensity trajectory at 100 ms time bins measured by a single photon counting module and **(B)** the histogram probability of different intensity bins divided into the "bright" (orange), "grey" (dark red), and "dark" (purple) intensity states typically seen in colloidal QDs. **(C)** Second order correlation function of photons in the HBT setup and their fits (dark red- with EF, grey- without) with G2 contrast of 0.01±0.05 for the 0V measurements and 0.05±0.05 for the +180V measurements. **(D)** "bright" state PL time trace, showing a bi-exponential fit to both +180V (red) and 0V (grey) time bins of the measurement. The multi exponential PL time trace at the "bright" state is a signature of the CQDM compared to the mono-exponential trace in the mono-QD (Fig. S7). **(E)** "grey" state PL time trace, showing a bi-exponential fit to both +180V (red) and 0V (grey) time bins of the measurement. **(F)** "dark" state PL time trace, showing a tri-exponential fit to both +180V (red) and 0V (grey) time bins of the measurement. **(G)** Fluorescence Intensity Lifetime distribution (FLID) of the CQDM for both voltage traces. Here the lifetime is calculated as the $I(\tau) = I_0 \cdot e^{-1}$ intensity reduction. The +180V (red) lifetime is elongated slightly for the "dark" states under 500 counts, and for the "bright" states above 1000 counts. Indicating a decrease in the short lifetime rate amplitude (leading to an overall longer time decay) and in the long rate time constant, for all three states, without changing the overall PL intensity.

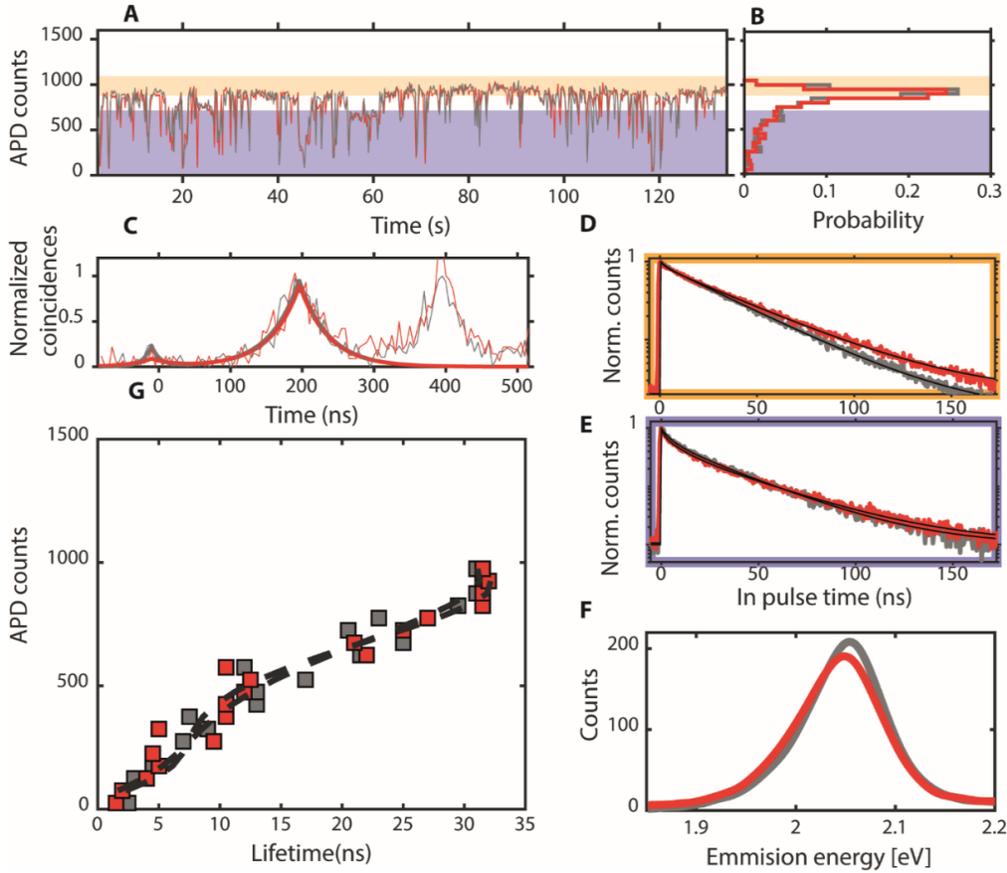

Figure S7. Example of typical emission shifting with applied EF in mono-QD. (A) Emission intensity trajectory at 100 ms time bins measured by a single photon counting module and (B) the histogram probability of different intensity bins divided into the "bright" (orange), and "dark" (purple) intensity states typically seen in colloidal QDs. (C) Second order correlation function of photons in the HBT setup (dark red- with EF, grey- without) with the G2 contrast calculated from the fit, showing 0.07±0.05 for both the +180V and 0V measurements. (D) "bright" state PL lifetime, showing a mono-exponential fit to both +180V (red) and 0V (grey) time bins of the measurement, which indicates the mono-QD. (E) "dark" state PL time trace, showing a tri-exponential fit to both voltage steps. (F) +180V (red) and 0V (grey) accumulated spectra from 1100-time frames of 100 ms each in the measurement, the peak difference between voltage traces is a 5±2 meV, where the +180V spectrum is red shifted. (G) Fluorescence Lifetime Intensity distribution (FLID) for both voltages, +180V (red) and 0V (grey), exhibiting similar trend.

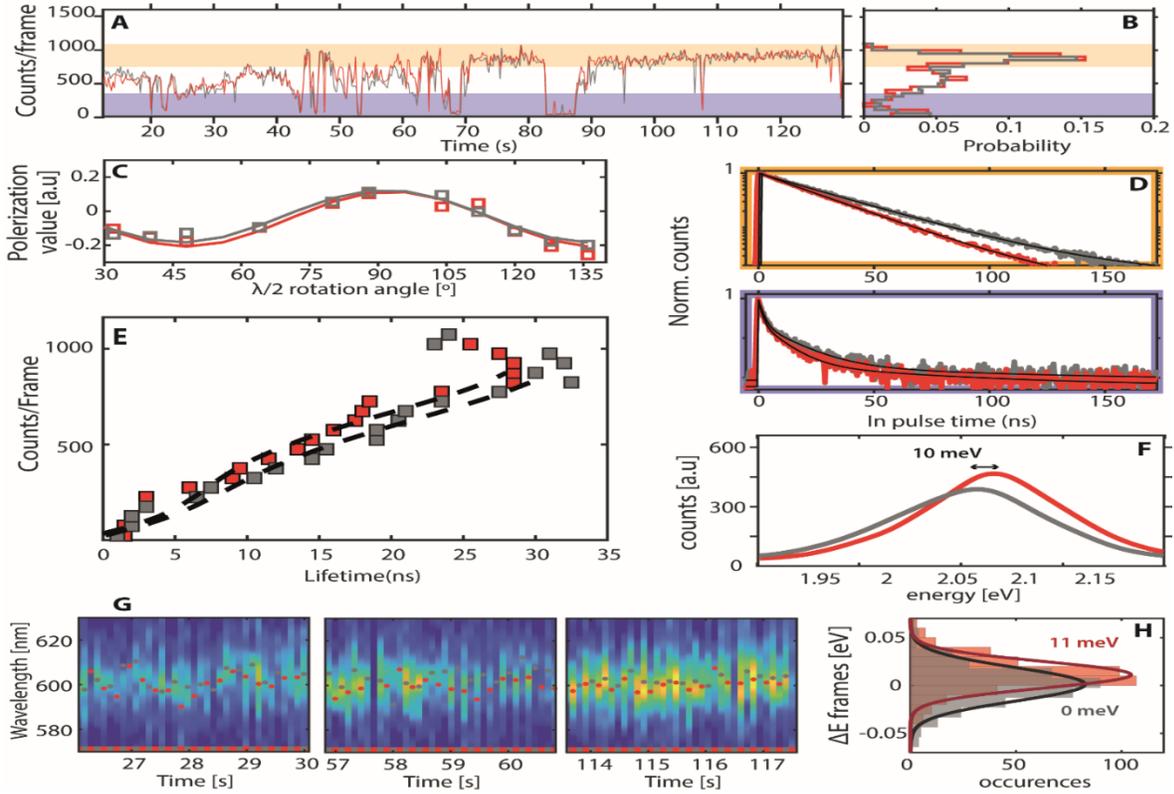

**Figure S8. Example of a maximal shifting mono-QD with C-axis polarized parallel to the applied electric field** (As explained in Figure S2). (**A**) Emission intensity trajectory and (**B**) the histogram probability of different intensity bins divided into the "bright" (orange), and "dark" (purple) intensity states. (**C**) PL polarization value fit for both voltage traces showing minor differences with a polarization angle of $\phi = 5 \pm 5°$ to the electrodes, indicating the QD's C-axis is perpendicular to them[4,5] and parallel to the EF (maximal effect). (**D**) "bright" state PL lifetime (orange panel), showing a mono-exponential fits and "dark" trace (purple panel) showing tri-exponential fits to both +180V (red) and 0V (grey) time bins of the measurement. (**E**) FLID for both voltage traces, exhibiting a similar trend (dashed lines) for both voltage traces. (**F**) +180V (red) and 0V (grey) accumulated spectra from 1100-time frames of 100 ms each in the measurement. The peak difference between voltage traces is a 10±2 meV, where the +180V is blue shifted. (**G**) Frame by frame time trace of the emission, with the gaussian fitted peak for each frame (red dots +180V, grey dots 0V) showing that spectral diffusion and charging events alter the magnitude and directionality of the EF induced emission shifting at different times in the measurement. (**H**) The mean frame shift analysis showing $< E_{frame}^{+180V} > = 11 \pm 2\ meV$ , and $< E_{frame}^{0V} > = 1 \pm 2\ meV$.

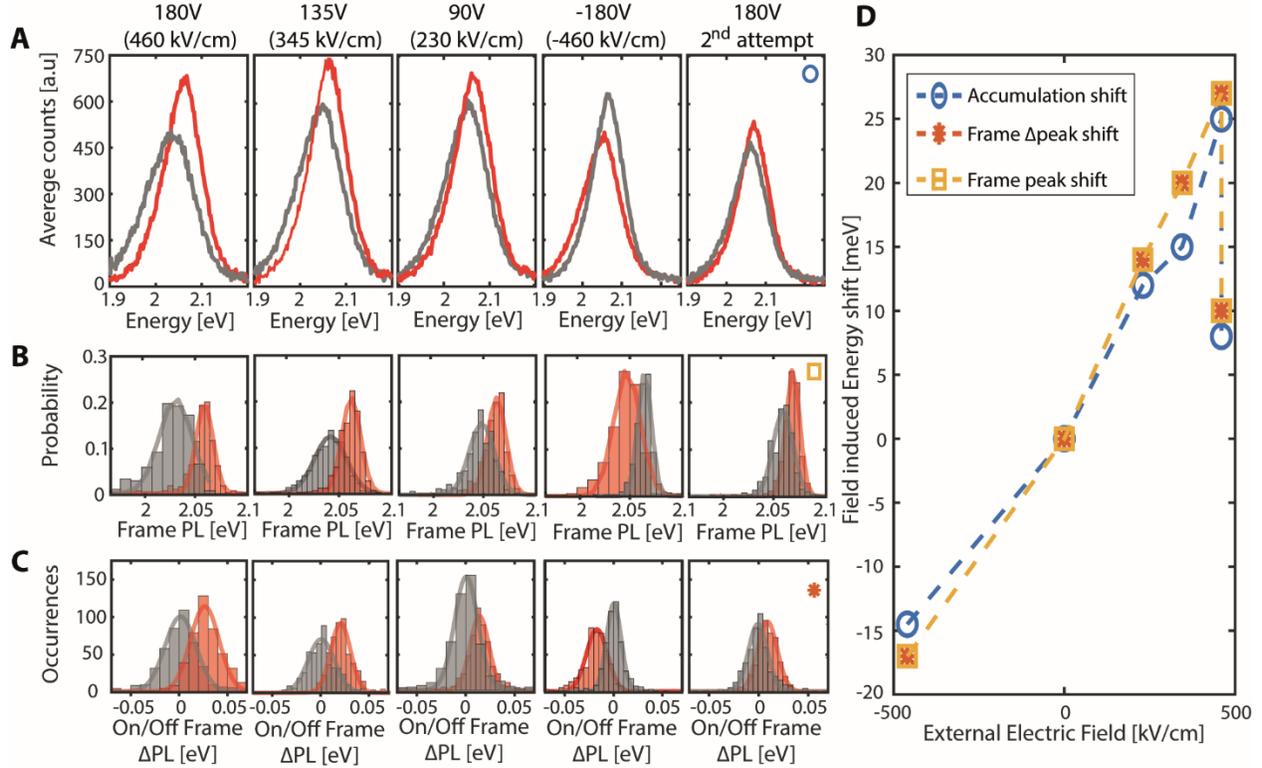

**Figure S9. Extended measurements of CQDM from Figure. 2 in main text at different applied EF conditions**. Accumulated emission energy spectra (raw data) (**A**), single time frame PL energy peak distribution (**B**), and calculated energy shift between successive with-without EF frames (**C**), of 1100-time frames of 100 ms each. For each measurement, a different voltage is applied on the electrodes, resulting in different induced EF (calculated by electrostatic finite element simulation in COMSOL Multiphysics, showed in figure S10). The measurement without applied EF (grey) shows narrowing and blue shifting of the accumulated (A) and average frame peak energy (B) with measurement progress (left to right), attributed to charging of the CQDM, while not affecting the spectral diffusion energy shifts without applied EF (C, grey). The measurement with applied EF (red) shows a consistent trend for the three analyses (**D**) in the energy shift regarding the EF polarity (positive/negative EF- blue/red shifts respectively). A second measurement in the same EF conditions as the first measurement (+180V, right-most graphs in A-C) gives a much weaker EF modulated effect which we link to the surface charging and dimming of the CQDM after over an hour of excitation.

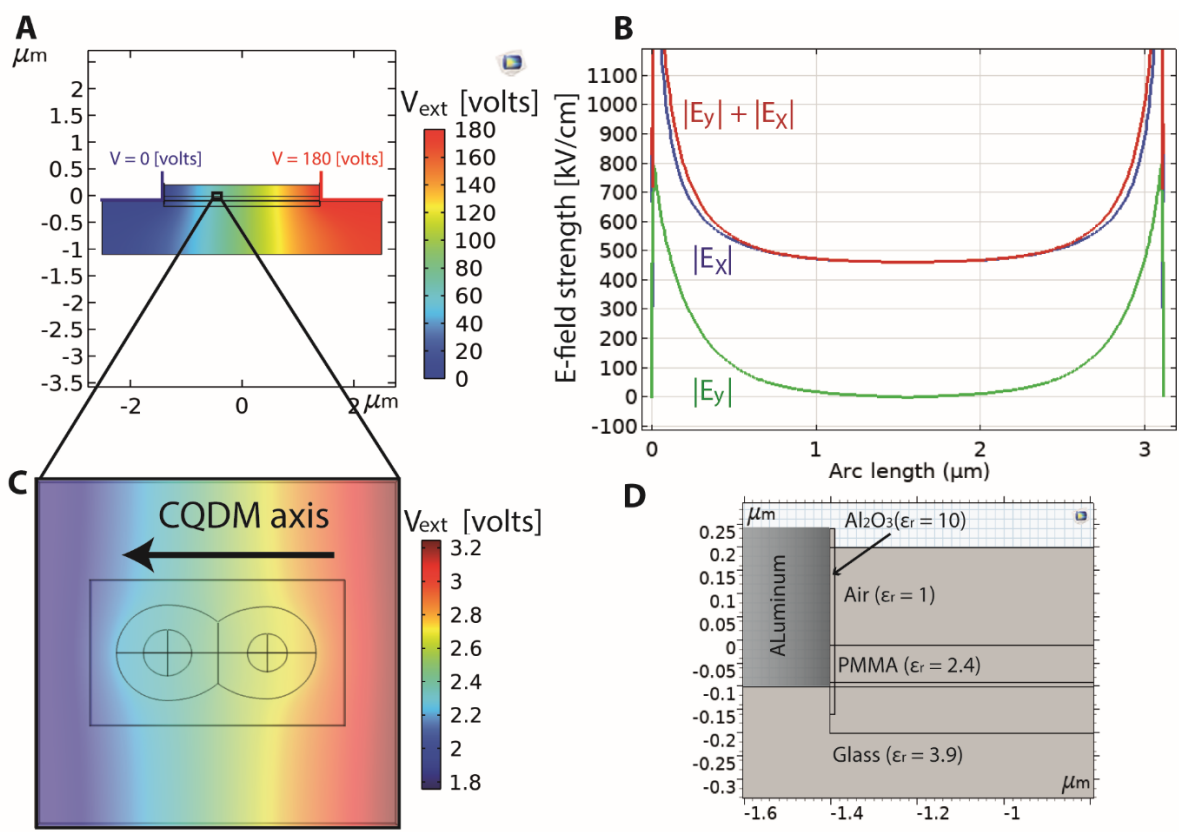

**Figure S10. Calculated external electric field between subsequent electrodes**. (**A**) The applied electric field is calculated by solving the Laplace equation on the electrode geometry, considering the dielectric polarization effects of the surroundings (glass substrate, PMMA, air). The result is an almost uniform field in the center (**B**) (nearly linearly changing voltage potential), parallel to the substrate plane between the electrodes. (**C**) The external electric potential on the CQDM is greatly affected by the dielectric screening that takes place in the interface between the nanocrystal surface and the PMMA surrounding. (**D**) the layered parameters of the relative dielectric constants of the different materials.

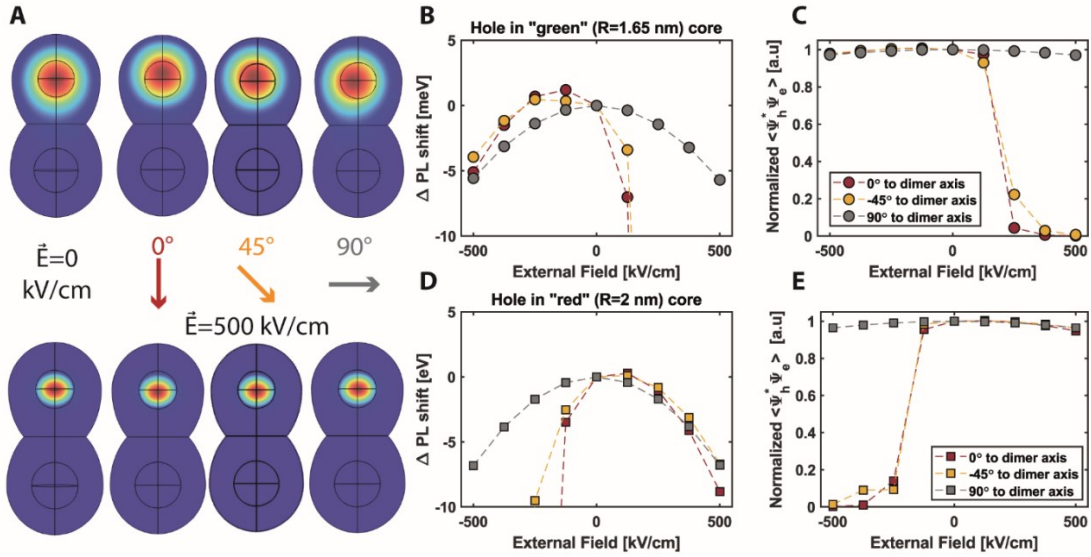

**Figure S11. Effect of Electric field on charge carrier wavefunction density.**

(**A**) simulation of exciton state (electron- top, hole- bottom) in the "yellow" QD of the CQDM, with/without applied electric field of 500 kV/cm at different angles to the fusion axis. Without the electric field, the hole is completely localized in the CdSe core due to the 0.64 eV valence band offset, while the electron is delocalized to the CdS shell, and also slightly shifted towards the second QD due to reduced confinement. With application of the electric field, the hole state remains confined in the CdSe core and is negligibly shifted in the different field directions, whereas the electron is strongly shifted in the opposite direction allowing tunability of its localization/delocalization between the two coupled QDs. (**B**) Exciton energy shift calculated for with - without electric fields at 0° (red), 45° (orange) and 90° (grey) relative to the fusion axis. (**C**) Electron-hole overlap normalized to the overlap value without EF of an exciton in the "yellow" emission center (66.7% for 0 applied electric field). (**D-E**) same as in B-C for an exciton in the "red" QD (71.7% for 0 applied electric field).

Normalized overlap under 20% is achieved for electric fields above 250 kV/cm (corresponding to application of above 90 Volts between 3 μm spaced electrodes), indicating a "switch-off" of one of the two emission centers by the field. Interestingly, in the opposite field polarity, up to -250 kV/cm applied for an exciton in the "yellow" center, the electric field induces localization to the center, resulting in slightly increased electron-hole overlap and up to 2 meV blue shift of the exciton energy. This phenomenon is increased for CQDMs with bigger size distribution (Figure S22).

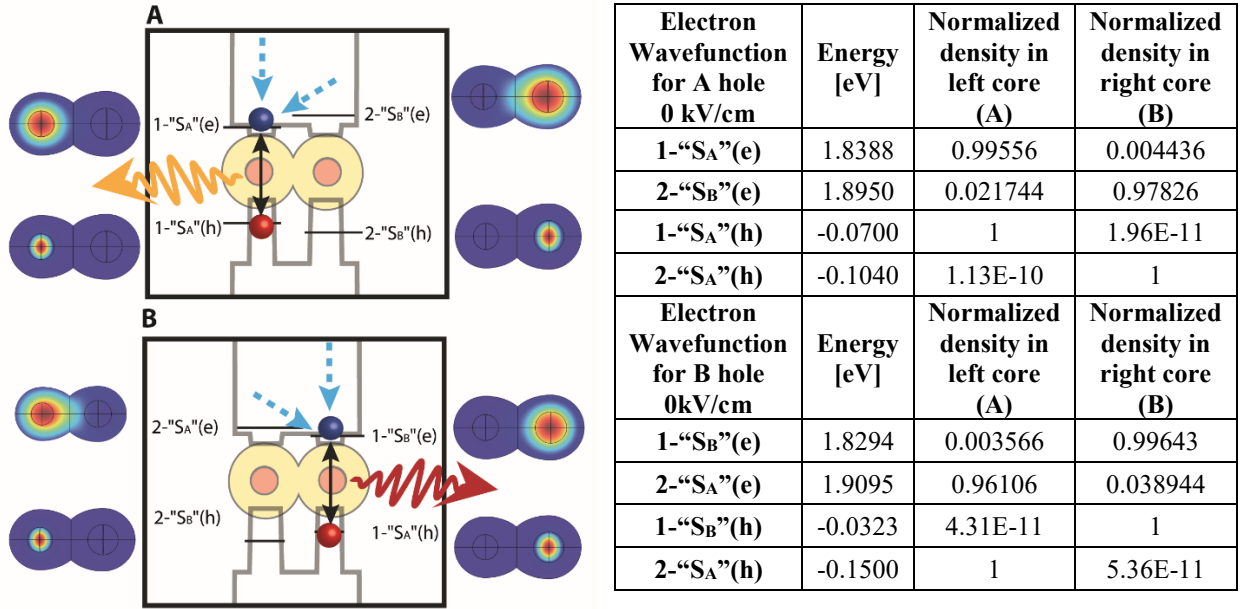

| Electron Wavefunction for A hole 0 kV/cm | Energy [eV] | Normalized density in left core (A) | Normalized density in right core (B) |
|---|---|---|---|
| **1-"S$_A$"(e)** | 1.8388 | 0.99556 | 0.004436 |
| **2-"S$_B$"(e)** | 1.8950 | 0.021744 | 0.97826 |
| **1-"S$_A$"(h)** | -0.0700 | 1 | 1.96E-11 |
| **2-"S$_A$"(h)** | -0.1040 | 1.13E-10 | 1 |
| **Electron Wavefunction for B hole 0kV/cm** | Energy [eV] | Normalized density in left core (A) | Normalized density in right core (B) |
| **1-"S$_B$"(e)** | 1.8294 | 0.003566 | 0.99643 |
| **2-"S$_A$"(e)** | 1.9095 | 0.96106 | 0.038944 |
| **1-"S$_B$"(h)** | -0.0323 | 4.31E-11 | 1 |
| **2-"S$_A$"(h)** | -0.1500 | 1 | 5.36E-11 |

**Figure S12. Calculated excited states for the CQDM in figure 1 without applied field**. (**A**) Schematic state diagram for the electron and hole wavefunctions without applied electric field. The Coulomb interaction is calculated for the electron with the hole state localized in the left (smaller) core, noted as the "S$_A$" state. (**B**) Same calculation with the Coulomb interaction now calculated for the electron with the hole state localized in the right (bigger) core, noted as the "S$_B$" state.

For both cases, it can be seen that the higher energetic state of the electron, localized in the left/right emission center, spills into the opposite QD rendering the higher state unstable, and therefore it relaxes to the lower state which is in the same QD as the hole due to the Coulomb interaction potential. The table shows calculated energies of the first and second electron and hole wavefunction states, and their normalized density in the yellow/red core (marked "A"/" B" respectively). States which have more than 1% density in the second emission center are considered delocalized and can toggle between the states.

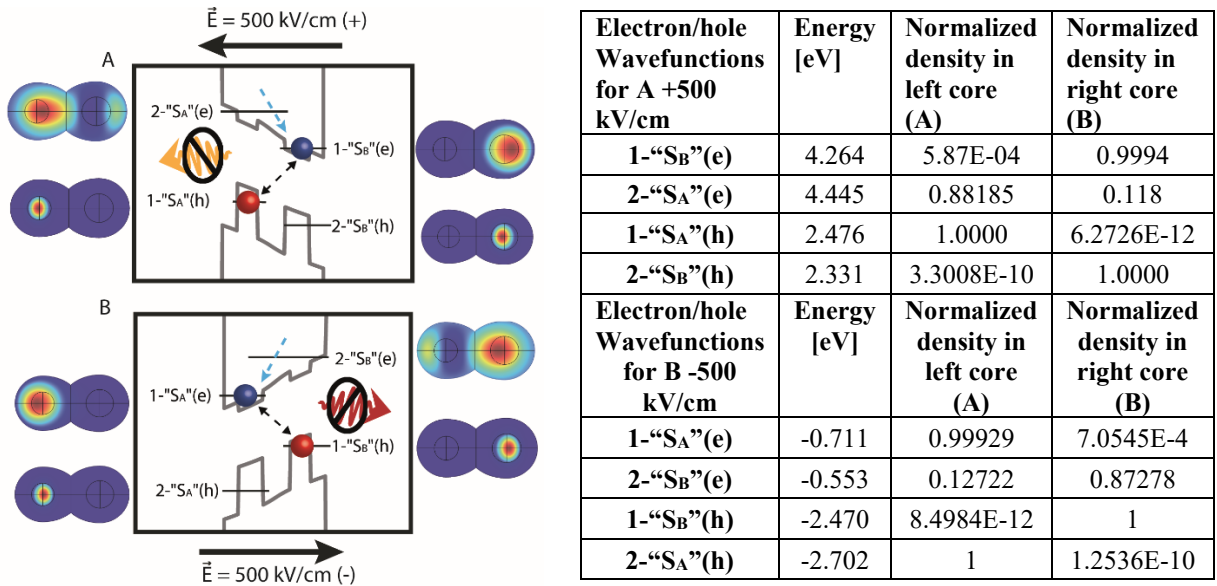

| Electron/hole Wavefunctions for A +500 kV/cm | Energy [eV] | Normalized density in left core (A) | Normalized density in right core (B) |
|---|---|---|---|
| 1-"$S_B$"(e) | 4.264 | 5.87E-04 | 0.9994 |
| 2-"$S_A$"(e) | 4.445 | 0.88185 | 0.118 |
| 1-"$S_A$"(h) | 2.476 | 1.0000 | 6.2726E-12 |
| 2-"$S_B$"(h) | 2.331 | 3.3008E-10 | 1.0000 |
| Electron/hole Wavefunctions for B -500 kV/cm | Energy [eV] | Normalized density in left core (A) | Normalized density in right core (B) |
| 1-"$S_A$"(e) | -0.711 | 0.99929 | 7.0545E-4 |
| 2-"$S_B$"(e) | -0.553 | 0.12722 | 0.87278 |
| 1-"$S_B$"(h) | -2.470 | 8.4984E-12 | 1 |
| 2-"$S_A$"(h) | -2.702 | 1 | 1.2536E-10 |

**Figure S13. Calculated excited states for the CQDM in figure 1 with 500 kV/cm applied field.** (**A**) state diagram for the electron and hole wavefunctions under high applied electric field. The Coulomb interaction is calculated for the electron with the hole state localized in the left (smaller) core, noted as the "$S_A$" state. The lowest and only localized state of the electron is in the right (bigger) core, due to the EF potential which overcomes the hole-electron Coulomb potential and solely dictates the electron location in the lowest radiative state ("$S_B$"). The second higher state of the electron ("$S_A$") manifests delocalization to the right core under the applied EF, and is probable to relax to the "$S_B$" state faster than the radiative recombination pathway with a hole in the left QD. (**B**) Same calculation with the Coulomb interaction now calculated for the electron with the hole state localized in the right (bigger) core, noted as the "$S_B$" state.

For both cases, the electron states are pushed to the emission center opposite to the field direction. The higher energetic state of the electron (2-"$S_B$"), is delocalized between the QDs and renderers the higher state unstable, and therefore it will relax to the lower state which is localized in the QD opposite the field direction, overcoming the Coulomb interaction potential between the charge carriers, and preventing radiative emission. The table shows calculated energies of the first and second electron and hole wavefunction states, and their normalized density in the yellow/red core (marked "A"/" B" respectively). States which have more than 1% density in the second emission center are considered delocalized and can toggle between the states.

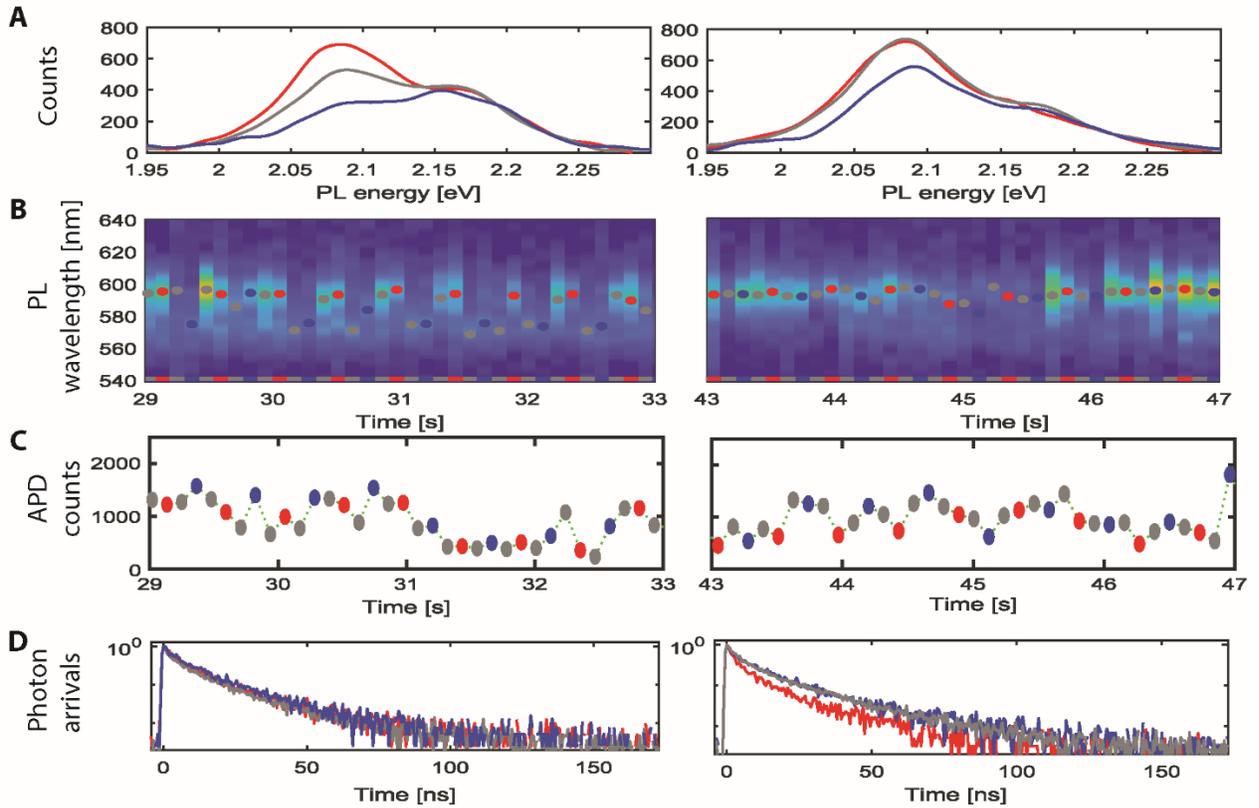

**Figure S14. Selected time traces from the measurement of the weakly fused hetero-CQDM in figure 4C.** (**A**) Accumulated 4 second emission spectrum of selected parts from the full 3-minutes measurement of the weakly-fused hetero-CQDM shown in Fig. 4C of the main text, showing an electric field induced emission dimming effect of the "red" emission center, effectively switching the emission from the "red" center to the "yellow" (left panels), or dimming the emission while still keeping the "red" peak the more dominant of the two (right panels). In all panels the different applied voltages are in a +180V(red), 0V(grey), -180V(blue), 0V(grey) sequence. (**B**) 100 ms time frames showing the emission and its gaussian fitted peak. Color switching is seen between the two peaks at 30-33 seconds (left panel), while dimming effect induces only small energy shifts (right panel). (**C**) photon counts measured on an Avalanche Photo-Diode (APD) showing no clear trend between the emission peak and emission intensity. (**D**) Emission photoluminescence lifetime showing almost no change between different applied potentials for the state with major PL color switching (left panel), in contradiction to the expected link between electron-hole overlap changes to energy shifting by the standard quantum confined stark effect in nanocrystalss manifesting a single emission center. The second "dimming" time-trace (right panel) shows shortening of the lifetime for the +180V time frames.

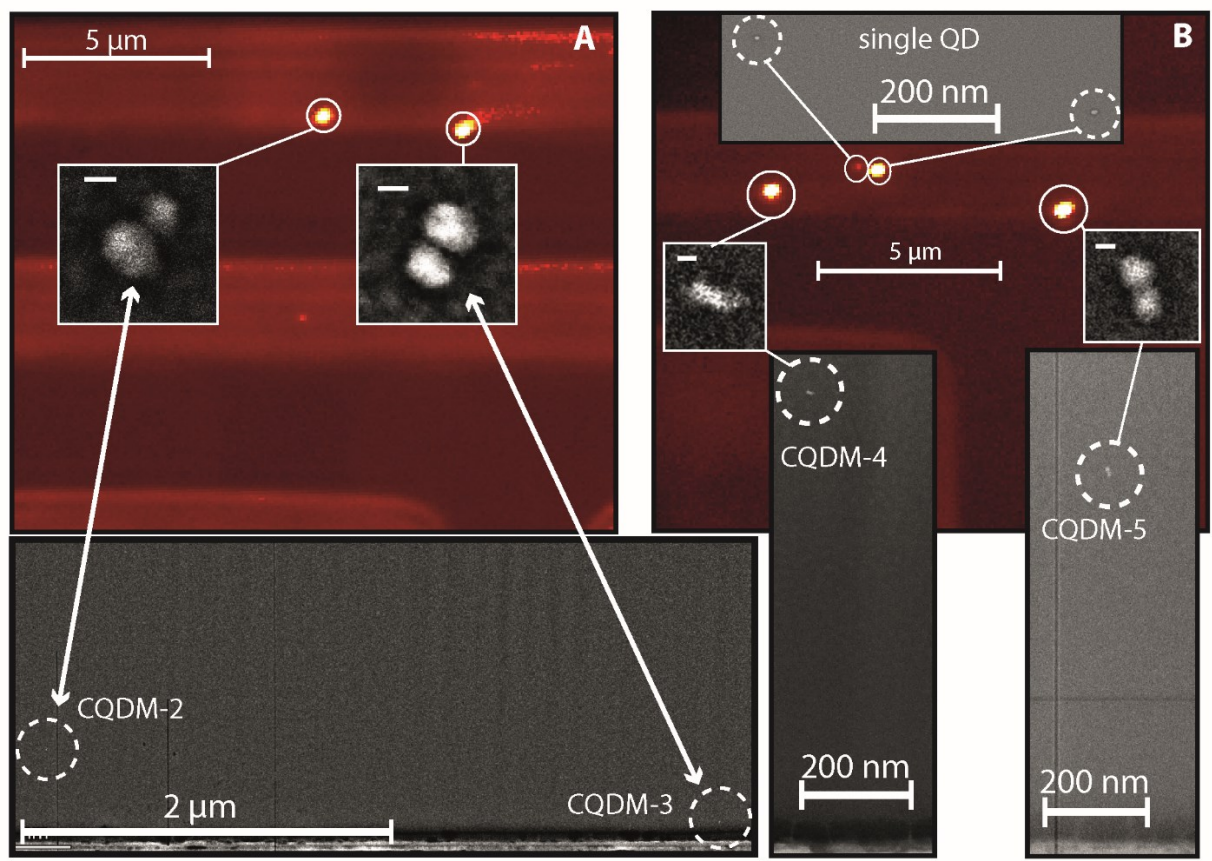

**Figure S15. More optical HAADF correlations of CQDMS.** (**A**) A widefield PL image of CQDMS showing a weakly fused (CQDM-2) and non-fused (CQDM-3) CQDMs. Insets show HAADF images at different magnification sizes of the lift out FIB-grid, showing direct spatial correlation between the widefield optical images of the CQDMS and their locations found in the scanning transmission electron microscope. (**B**) An additional widefield PL image area with 2 CQDMS (4,5, sides) and 2 additional single QDs (center). Scale bars of all HAADF close-up images (CQDMs 2-5) are 5 nm.

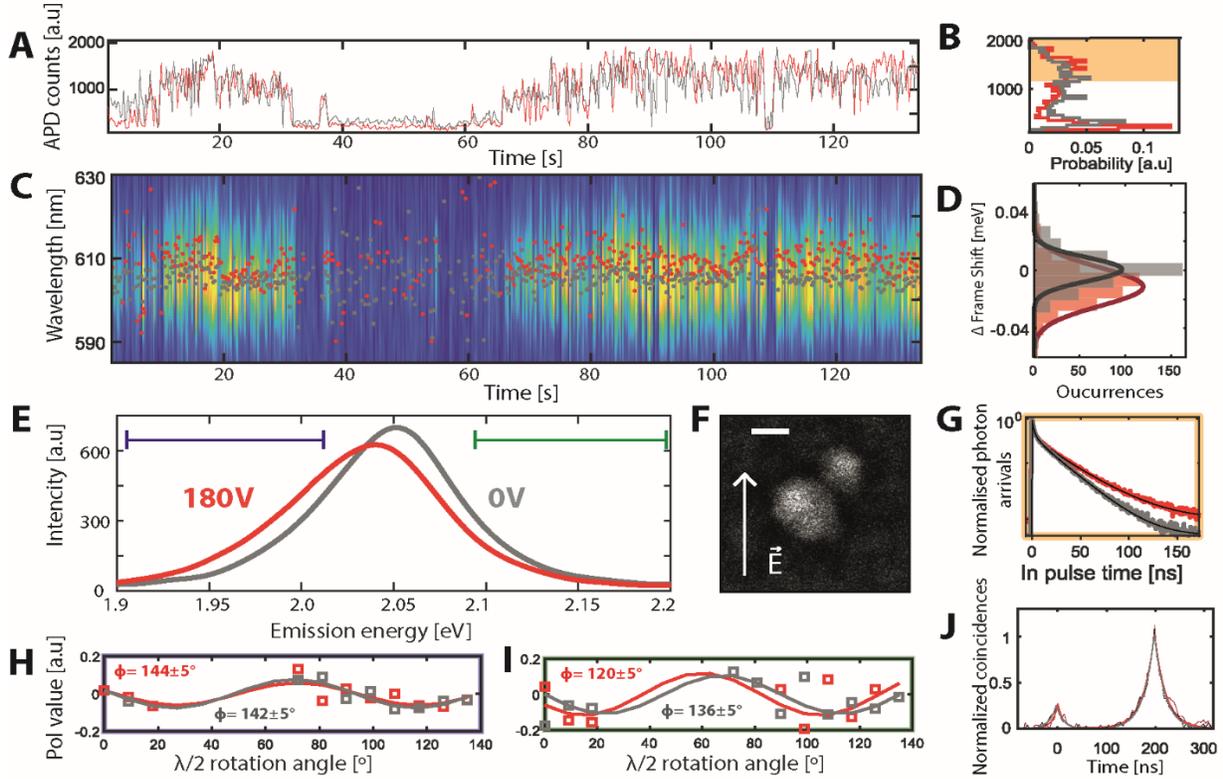

**Figure S16. Effect of EF on Emission spectra and photon statistics for CQDM-2.** (**A**) Emission intensity trajectory at 100 ms time bins measured by a single photon counting module and (**B**) the histogram probability of different intensity bins, with the "bright" intensity state marked in orange. (**C**) frame by frame spectra showing that the color switching is not robust throughout the measurement. The mean frame shift analysis (**D**) shows that the electric field induced shifts (red) are broader than the spectral diffusion. The accumulated PL spectra (**E**) does not show two distinguishable emission peaks as is expected for the size difference (~3 nm diameter difference) seen in the HAADF image, which also shows low neck filling between the dots (**F**). This may indicate the color switching is not between the color centers but between a confined state on the bigger QD and a de-confined state in the interface of the two dots. (**G**) "bright" state PL time trace, showing a tri-exponential fit to both +180V (red) and 0V (grey) time bins of the measurement, indicating that charged states are dominant in the emission. The emission polarization degree (**H-I**) barely shifts between the low (purple-H) to high (green-I) energy photons from the edges of the emission spectrum, marked in (**E**). (**J**) Second order correlation function of photons in the HBT setup (dark red- with EF, grey- without) with the g²(0) contrast calculated from the fit, showing 0.14±0.04 for the 0V measurements and 0.17±0.05 for the +180V measurements, indicating low biexciton quantum yield despite the observed weak fusion between the two QDs.

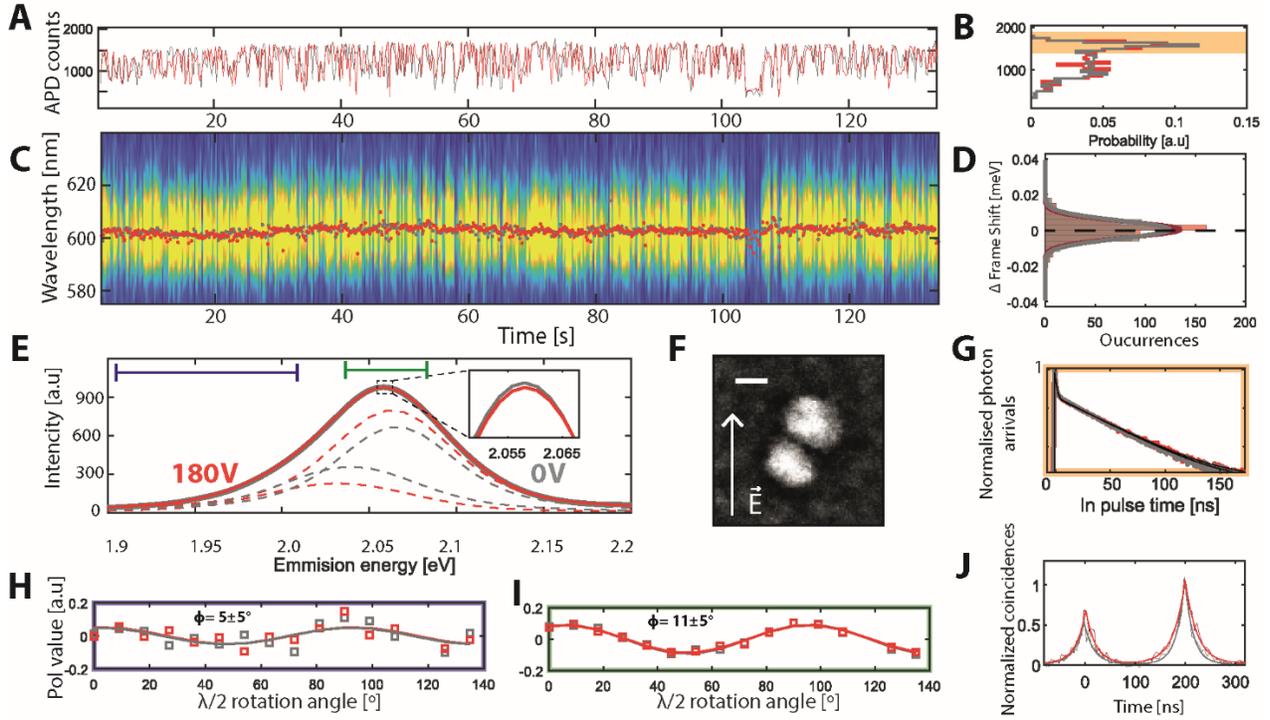

**Figure S17. Effect of EF on emission spectra and photon statistics for CQDM-3.** (**A**) Emission intensity trajectory at 100 ms time bins measured by a single photon counting module and (**B**) the histogram probability of different intensity bins, with the "bright" intensity state marked in orange. (**C**) frame by frame spectra showing no noticeable color switching throughout the measurement. The mean frame shift analysis (**D**) also shows that the electric field induced shifts (red) are equivalent to the spectral diffusion. The accumulated PL spectra (**E**) does not show two distinguishable emission peaks or shifting emission peaks. HAADF image (**F**) shows two QDs with approximately 1 nm spacing between them, which prevents electron shifting between dots and only long-distance energy transfer mechanisms may take place in this system. (**G**) "bright" state PL time trace, showing a bi-exponential fit to both +180V (red) and 0V (grey) time bins of the measurement. The emission polarization degree (**H-I**) barely shifts between the low (purple-H) to mid (green-I) energy photons (high energy photons did not give a good fit) from the emission spectrum, which may indicate that the two QDs are aligned in the emission dipole. (**J**) Second order correlation function of photons in the HBT setup (dark red- with EF, grey- without) with the g²(0) contrast calculated from the fit, showing 0.53±0.06 for the 0V measurements and 0.49±0.05 for the +180V measurements, indicating high biexciton quantum yield which comes mostly from segregated biexcitons.

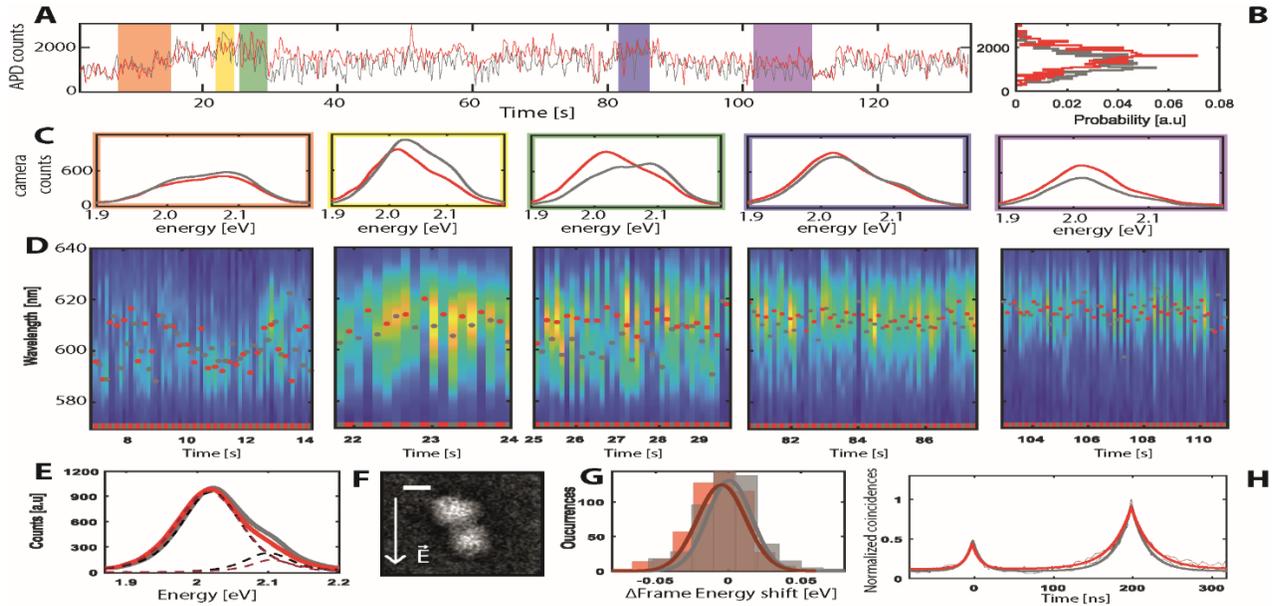

**Figure S18. Effect of EF on emission spectra and photon statistics for CQDM-5.** (**A**) Emission intensity trajectory at 100 ms time bins measured by a single photon counting module and (**B**) the histogram probability of different intensity bins. The accumulated (**C**) and frame-by-frame (**D**) emission spectra show that the CQDM is affected differently by the applied electric field at different times of the measurement. Dimmed emission is hardly affected (orange), bright states can be dimmed by the applied field (yellow) or exhibit partial color switching (green) between the ~90 meV spaced emission centers. Otherwise, the switching can be weak (purple) and dimmed emission traces can be enhanced (light purple). The accumulated spectra from the entire measurement (**E**) shows an overall effect of dimming of the higher energy emission center under applied EF. HAADF image (**F**) shows a CQDM aligned at 20° to the applied field, with the bigger QD opposite to the field direction, explaining the partial switching off of the smaller QD. Mean frame shifts (**G**) of the emission are slightly red shifted with the applied field, (**H**) while the second order photon coincident counts show $g^2(0)$ contrast of 0.25±0.08 for the 0V measurements and 0.16±0.05 for the +180V measurements, indicating higher than mono-QD biexciton quantum yield, emitted from both segregated and non-segregated biexcitons. High biexciton to exciton quantum yield ratio can explain the "not full" color switching seen for this CQDM, resulting from a high barrier between the fused QDs that causes relatively weak electronic coupling.

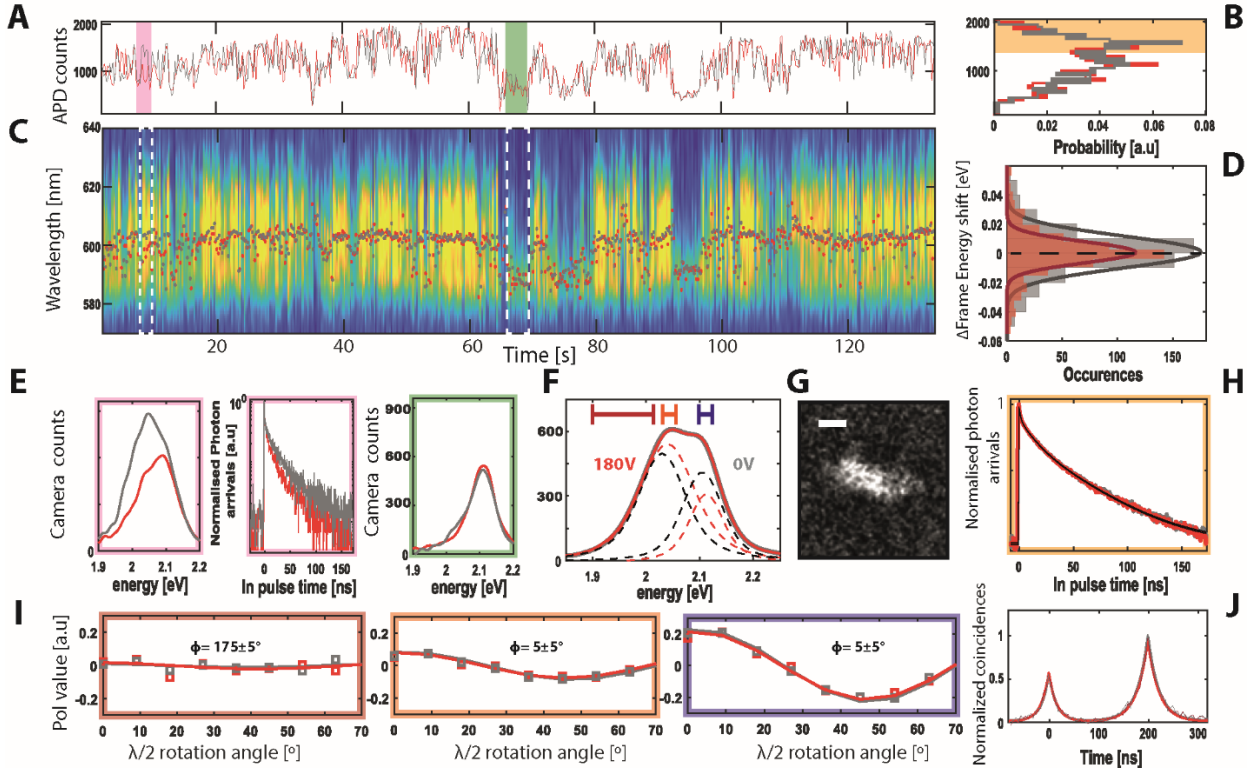

**Figure S19. Effect of EF on emission spectra and photon statistics for CQDM-4.** (**A**) Emission intensity trajectory at 100 ms time bins measured by a single photon counting module and (**B**) the histogram probability of different intensity bins, with the "bright" intensity state marked in orange. (**C**) frame by frame spectra showing no noticeable color switching throughout the measurement, and emission dimming of the "red" emission center for short periods in the time trace. The mean frame shift analysis (**D**) also shows that the electric field induced shifts (red) are equivalent to the spectral diffusion shifts. (**E**) Within the measurement, an example of emission dimming and switching between the "red" to "yellow" emission centers is shown (pink boxes), accompanied with radiative lifetime shortening and intensity dimming. For an in-measurement time where emission is seen only from the "yellow" center (green box) minor effects of the electric field on the emission are observed. The accumulated emission spectra from the entire measurement (**F**) shows two distinguishable emission peaks separated by ~75 meV, and no distinguishable changes are seen with application of the external electric field. HAADF image, though of low resolution, (**F**) shows an elongated CQDM aligned almost perpendicular to the applied field (from top to bottom of the image). "Bright" state PL time trace (**H**), showing a tri-exponential fit to both +180V (red) and 0V (grey) time bins of the measurement indicating that charged states also emit. The emission polarization degree (**I**) is nearly the same in three indicative parts of the spectrum, showing the two QDs are fused with their C axis at the same direction to each other, and in the direction of the applied field. (**J**) Second order correlation function of photons in the HBT setup (dark red- with EF, grey- without) with the $g^2(0)$ contrast calculated from the fit, showing 0.36±0.04 for the 0V measurements and 0.42±0.05 for the +180V measurements, indicating high biexciton quantum yield which comes mostly from segregated biexcitons. High biexciton quantum yield and negligible electric field effect contradict the full fusion neck filling seen in the HAADF image. This may indicate that even with fusion, stacking faults and defects in the lattice filling between the two QDs can maintain a high energy barrier between them, preventing electron delocalization between dots and a robust color switching electric field effect. In addition, the 75° angle between the CQDM axis and the electric field direction results in a low effective field in the axis direction, as seen also in the simulation in figure 3.

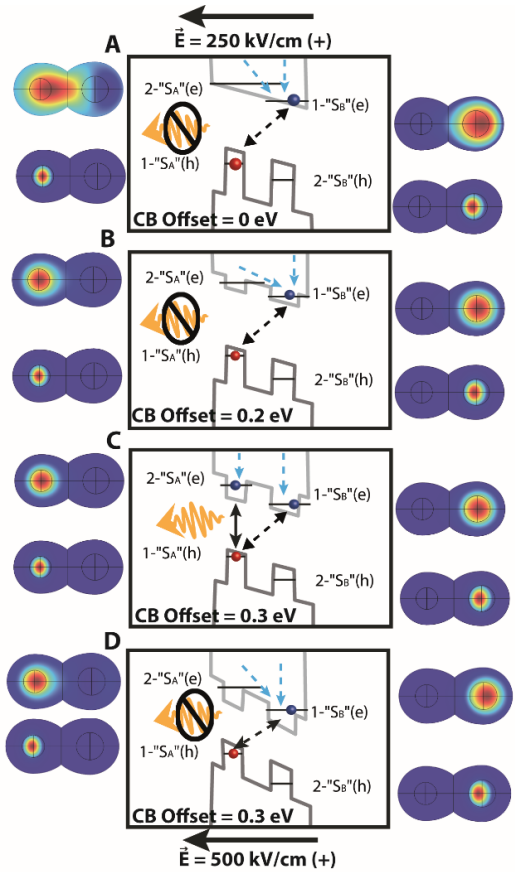

| Electron Wavefunction for A 250 kV/cm | Energy [eV] | Normalized density in left core (A) | Normalized density in right core (B) |
|---|---|---|---|
| 1-"S_B"(e) | 3.013 | 0.006724 | 0.99328 |
| 2-" S_A"((e) | 3.080 | 0.8063 | 0.1937 |
| Electron Wavefunction for B 250 kV/cm | Energy [eV] | Normalized density in left core (A) | Normalized density in right core (B) |
| 1-"S_B"(e) | 3.119 | 7.5688E-4 | 0.99924 |
| 2-" S_A"((e) | 3.220 | 0.98964 | 0.010356 |
| Electron Wavefunction for C 250 kV/cm | Energy [eV] | Normalized density in left core (A) | Normalized density in right core (B) |
| 1-"S_B"(e) | 3.160 | 2.7216E-4 | 0.99973 |
| 2-" S_A"((e) | 3.272 | 0.99800 | 0.0020036 |
| Electron Wavefunction for D 500 kV/cm | Energy [eV] | Normalized density in left core (A) | Normalized density in right core (B) |
| 1-"S_B"(e) | 4.327 | 7.8458E-5 | 0.99992 |
| 2-" S_A"((e) | 4.594 | 0.98654 | 0.013462 |

**Figure. S20. Calculated excited states for the CQDM in figure 1 for different simulated conduction band offsets at representative fields.** (**A**) Calculation for 0 eV conduction band (CB) offset at 250 kV/cm applied electric field. The higher excited electron state is delocalized in both QDs, relaxing to the lower localized excited state in the "red" QD, preventing emission from the "yellow" QD. (**B**) For a 0.2 eV CB offset at 250 kV/cm, the higher excited state is much less delocalized but still may relax to the lower energy QD. (**C**) For a higher CB offset of 0.3 eV, at 250 kV/cm the first and second electron excited states are localized in each QD separately, preventing relaxation between them, and allowing emission from both cores. Only at strong applied fields of 500 kV/cm (**D**) does the higher excited electron state "spill" out to the "red" QD allowing relaxation and turning of the emission pathway from the "yellow" core. The table shows calculated energies of the first and second electron wavefunction states, and their normalized density in the yellow/red core (marked "A"/" B" respectively).

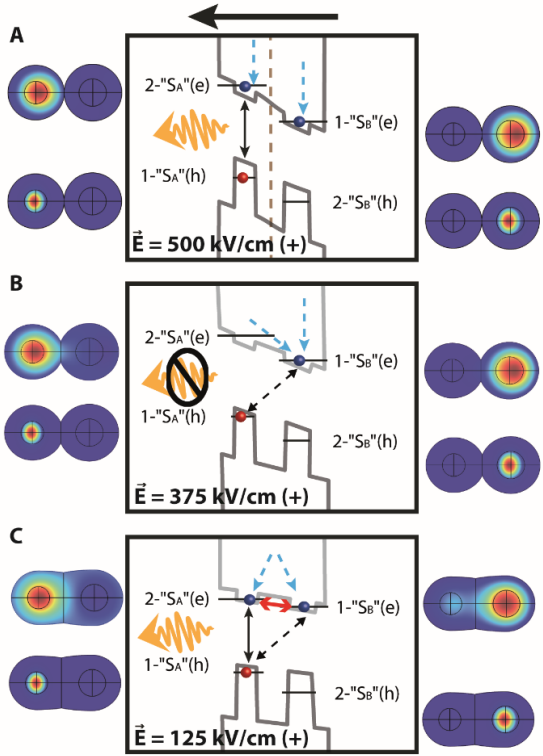

| Electron Wavefunction for A 500 kV/cm | Energy [eV] | Normalized density in left core (A) | Normalized density in right core (B) |
|---|---|---|---|
| 1-"S_B"(e) | 4.257 | 7.04E-06 | 0.99999 |
| 2-"S_A"(e) | 4.465 | 0.99973 | 2.67E-04 |
| **Electron Wavefunction for B 375 kV/cm** | **Energy [eV]** | **Normalized density in left core (A)** | **Normalized density in right core (B)** |
| 1-"S_B"(e) | 3.681 | 3.26E-04 | 0.99967 |
| 2-"S_A"(e) | 3.796 | 9.84E-01 | 1.64E-02 |
| **Electron Wavefunction for C 125 kV/cm** | **Energy [eV]** | **Normalized density in left core (A)** | **Normalized density in right core (B)** |
| 1-"S_A"(e) | 2.483 | 0.079946 | 0.92005 |
| 2-"S_B"(e) | 2.496 | 0.91177 | 0.088229 |

**Figure S21. Calculated excited states for the CQDM for different fusion neck widths at representative fields.** Calculations for (**A**) the "touching" CQDM (neck width 0.6 nm), showing localized electronic states in each QD separately even at 500 kV/cm applied electric field; (**B**) a CQDM with a 3 nm neck, in which delocalization of the higher electronic state "spilling" into the lower state QD is achieved at applied fields higher than 375 kV/cm; and (**C**) the "fully fused" CQDM (neck width of 7.5 nm), which exhibits electronic state spilling even at low applied electric fields of 125 kV/cm. The electronic state energetic difference between the two emission centers is reduced to less than the thermal energy at room temperature, leading to thermal state population of both excited states, where the higher exited state can radiatively recombine and thus the emission from the "yellow" center is still active. At higher applied electric fields this energy difference is increased above thermal mixing, resulting in relaxation of the higher state (in the yellow core) to the lower state and switching off the "yellow" emission (Fig. 5). The table shows calculated energies of the first and second electron wavefunction states, and their normalized density in the yellow/red core (marked "A"/" B" respectively).

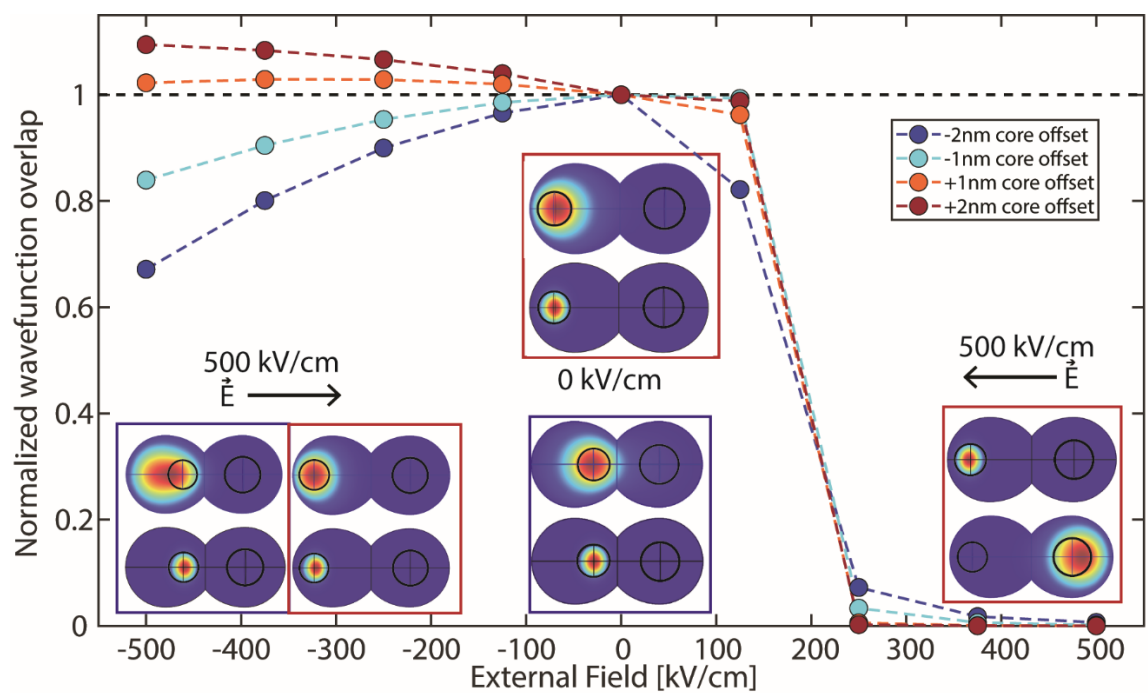

**Figure S22. Calculated normalized electron-hole overlap for the CQDM in figure 1 with different core locations.** The CQDM dimensions are 1.65 nm/2.8 nm (2 nm/2.45 nm) core-shell radii for the left (right) emission center, giving exciton energies of 2.030 and 1.990 eV and normalized electron-hole wavefunction overlap of 65% and 71% (left-right, respectively) with 0 kV/cm applied EF (insets, center). Introducing offsets in the smaller core location (left) in the range of ±2 nm from the QD center shows that when the core is closer to the second dot the electron wavefunction delocalizes more between the two dots at moderate fields. At high fields, the introduced offset alters the overlap on the same sided exciton between 110% to 67% (+2nm, -2nm) compared to the initial electron-hole overlap without EF (65% with ±1% difference for core location change). The change in the charge carrier overlap should result in change in the quantum yield of the emission center, similar to effects observed in similar elongated nanocrystals[2,3]. Thus, we may see an effect of emission dimming or enhancement on the emission center in one side of the CQDM, uncorrelated with the emission shifting on the other side. This effect is expected to be seen more clearly in cases where the intra-particle fusion is weak (as the two cores are less coupled), such as the CQDM example in figure 4C in the main text, where the lower energy emission center exhibits nearly 50% emission intensity difference between applied electric fields of ±180V.

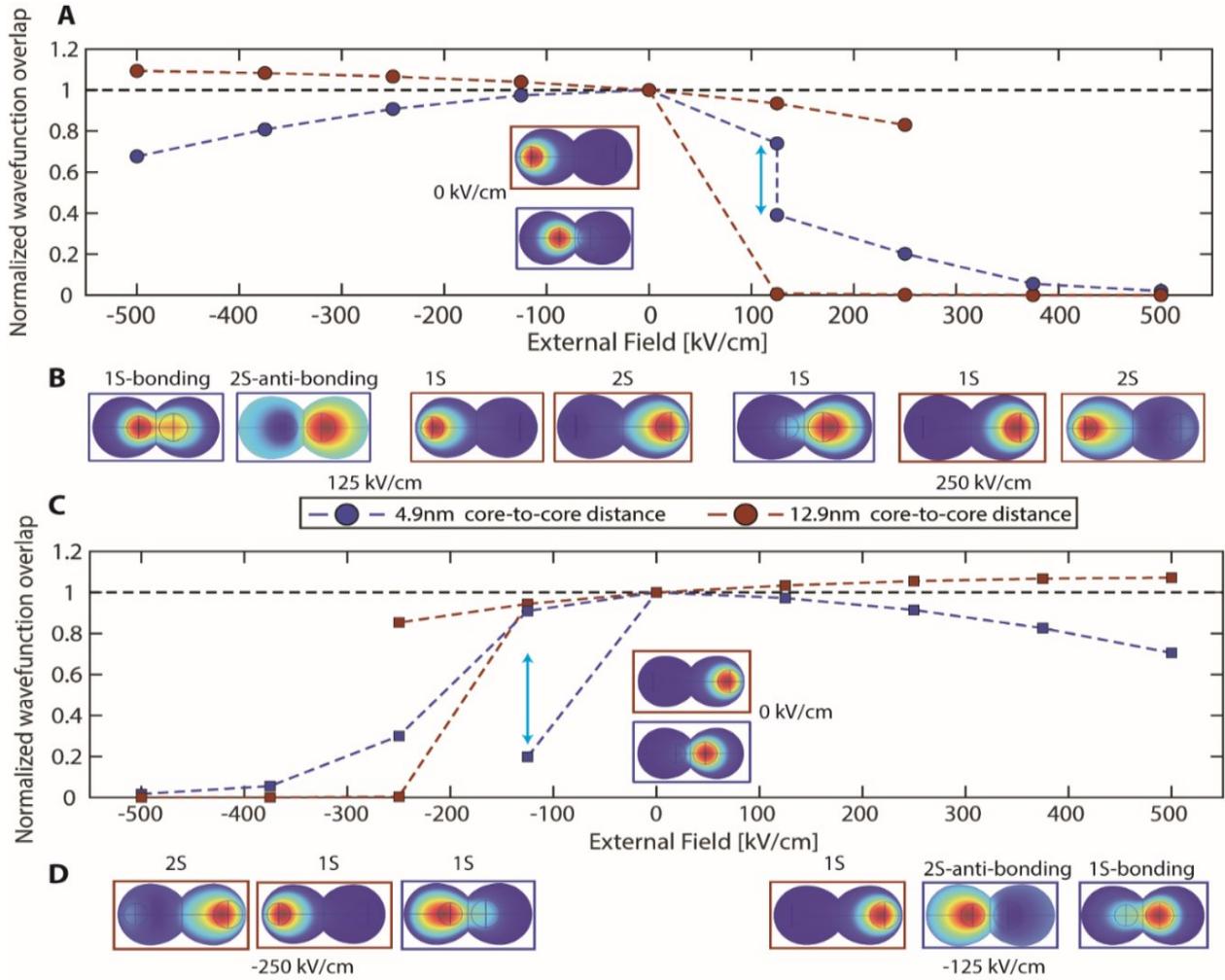

**Figure S23. Calculated normalized electron-hole overlap for the CQDM in figure 1 with different core-to-core distancing.** Using the same dimensions as in Fig. S22 and introducing offsets to both of the core locations in the range of ±2nm, we show the normalized electron-hole overlap integral. The simulated geometries exhibit a core-to-core center distance of 4.9 nm (blue) and 12.9 nm (dark red) as the extreme cases of core location offset in a CQDM. (**A, C**, inset shows the electron wavefunctions at 0 kV/cm applied EF) the normalized charge-carriers overlap for the case of the hole confined in the left (right) emission center. (**B, D**) electron wavefunction density of the lowest (1S) and second (2S) electron states in the CQDM calculation at moderate EFs. At ±125 kV/cm applied EF the electron states are hybridized between cores for the 4.9 nm distance, and also at ±250 kV/cm the electron is slightly delocalized between the sides, which should reduce the emission switching probability and may allow radiative recombination of an indirect exciton. The 12.9 nm separated cores show stable localized electron states at applied EFs up to 250 kV/cm, and the effect shows a similar trend as the increased potential barrier effect between sides showed previously in simulations changing the fusion-neck width and valence-conduction band offset between the CdSe-CdS heterostructure.

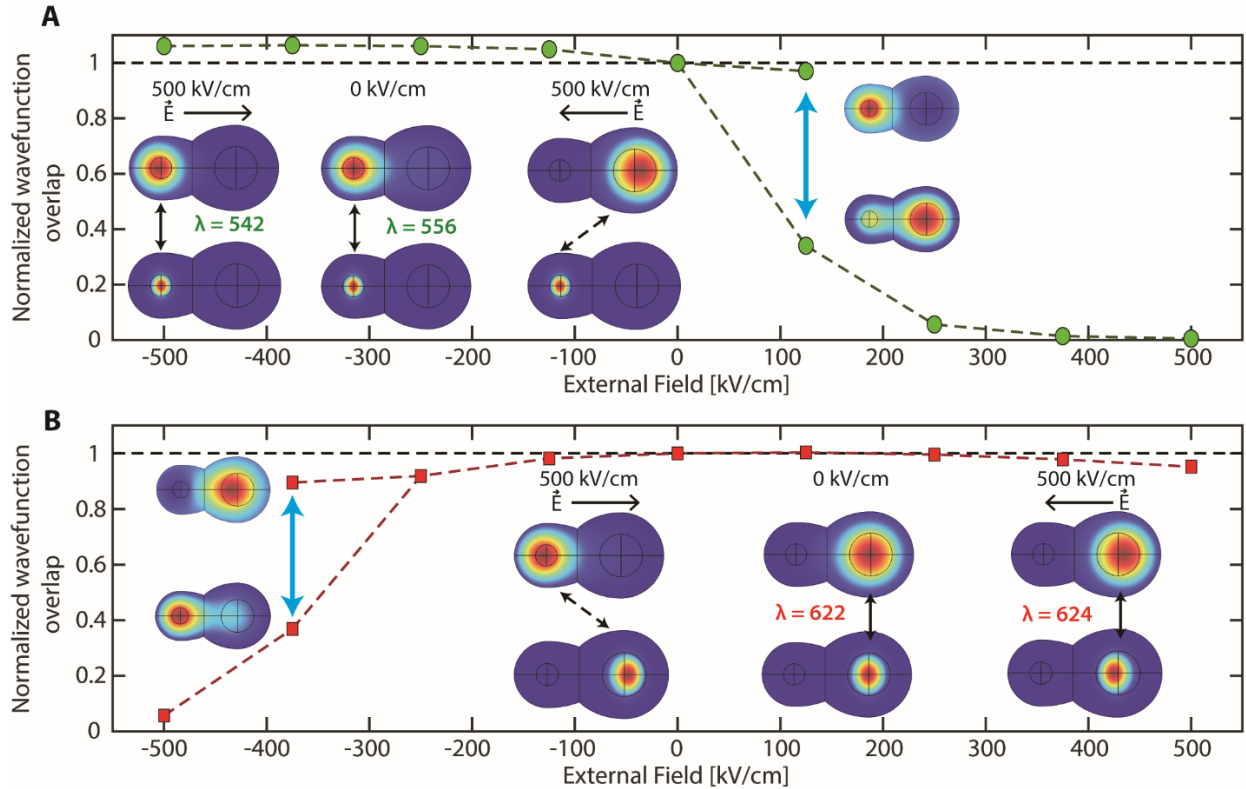

**Figure S24. Simulation of a Hetero-CQDM with red-green emission wavelengths.** The CQDM consists of two QDs with CdSe cores of 2:1 size (1.05 nm – green, 2.1 nm – red), coated by a 2.1 nm CdS shell and fused with a neck width of 5.24 nm. Without applied electric field, the exciton energy difference between the two QDs is 235 meV. The normalized electron-hole wavefunction overlap is calculated using the same procedure as in figure 3 (with a 0V value of 60/73% for the green/red emission center, respectively). Electron/hole wavefunctions are shown at different representative applied electric fields (inset). Electronic states where the hole is localized in the green core (**A**) are still "spilling" to the bigger QD, allowing color switching above 125 kV/cm. Applied electric field for an exciton in the green center can also induce electron localization in that QD, increasing the electron-hole overlap by up to 6.4% (-375 kV/cm), and blue shifting the emission by 58 meV. Switching the electron "up-hill" from red to the green (**B**) is more challenging, and a localized electron state in the green QD while the hole is still in the red core is seen onely at a -500 kV/cm applied field. For both red and green switching cases, we see a transition state in which the electron energy states are spaced apart by less than KT at room temperture (green: 125 kV/cm, red: -375 kV/cm) where thermal population of both states may prevent the electric field switching effect.

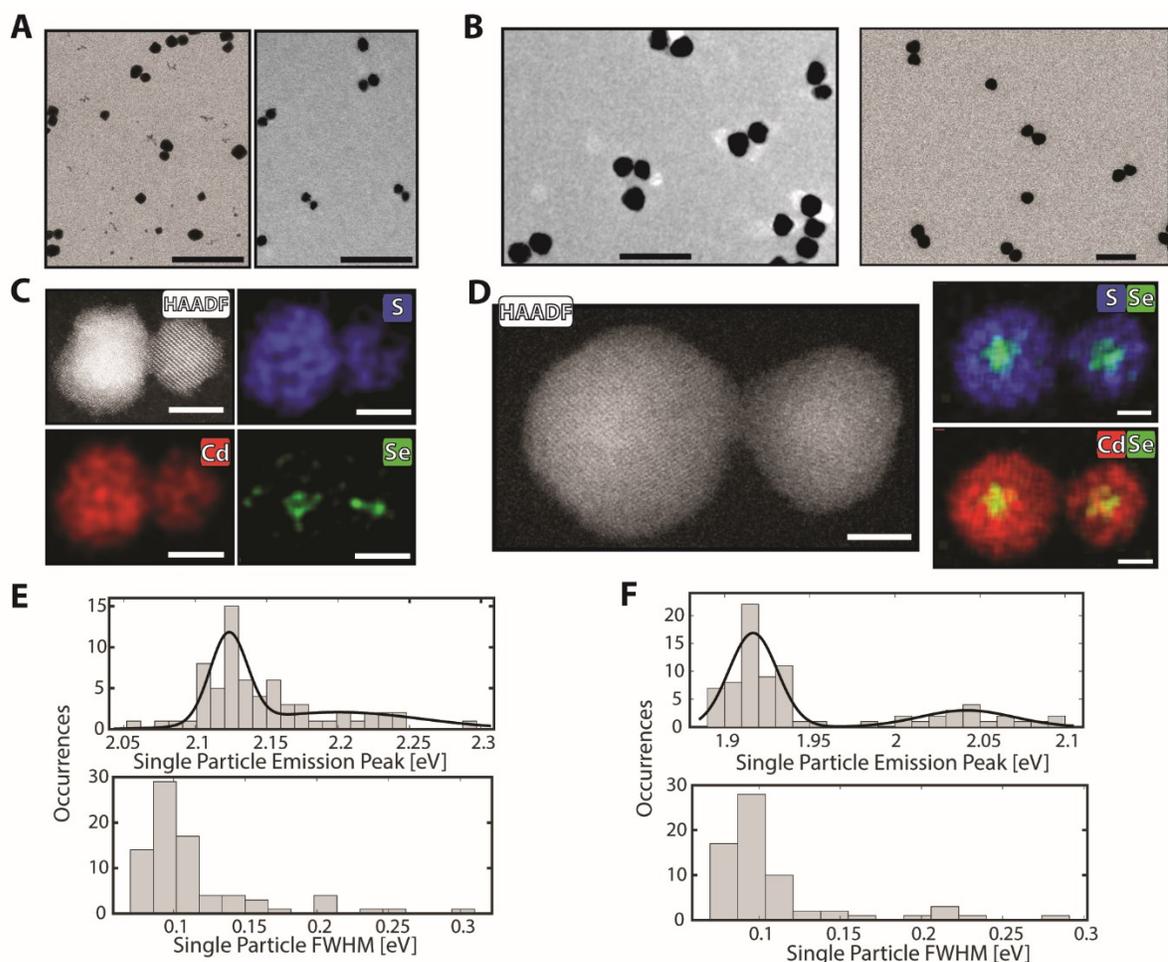

**Figure S25. Statistical data of hetero-sized CdSe/CdS CQDMs.** Left panels present the characterization of a CQDM sample made from 1.40±0.35 nm/2.1±0.7 nm core/shell QDs, and right panels present the characterization of a CQDM sample made from two distinct sizes of large QDs (2.2 nm/5.3 nm and 1.2 nm/4.8 nm core/shell QDs). (**A, B**) Transmission electron microscope images, showing that the CQDMs in B are roughly double the size of the CQDMs in A. Black scale bars are 50 nm. (**C, D**) HAADF and EDS elemental analysis for an example of a statistically created hetero-sized CQDM from each batch. White scale bars are 5 nm. (**E, F**) top panels are emission spectra peaks histograms for single particles from each sample, both showing two emission energy populations. The Full width at half maxima (FWHM) of the single particles are shown in the bottom panels of E-F for each sample. As mono-QDs of CdSe/CdS synthesized by this procedure[6] regularly exhibit emission FWHM of less than 100 meV, we can assume that the particles showing a FWHM of more than 150 meV are CQDMs from QDs with emission energy difference of above ~75 meV.

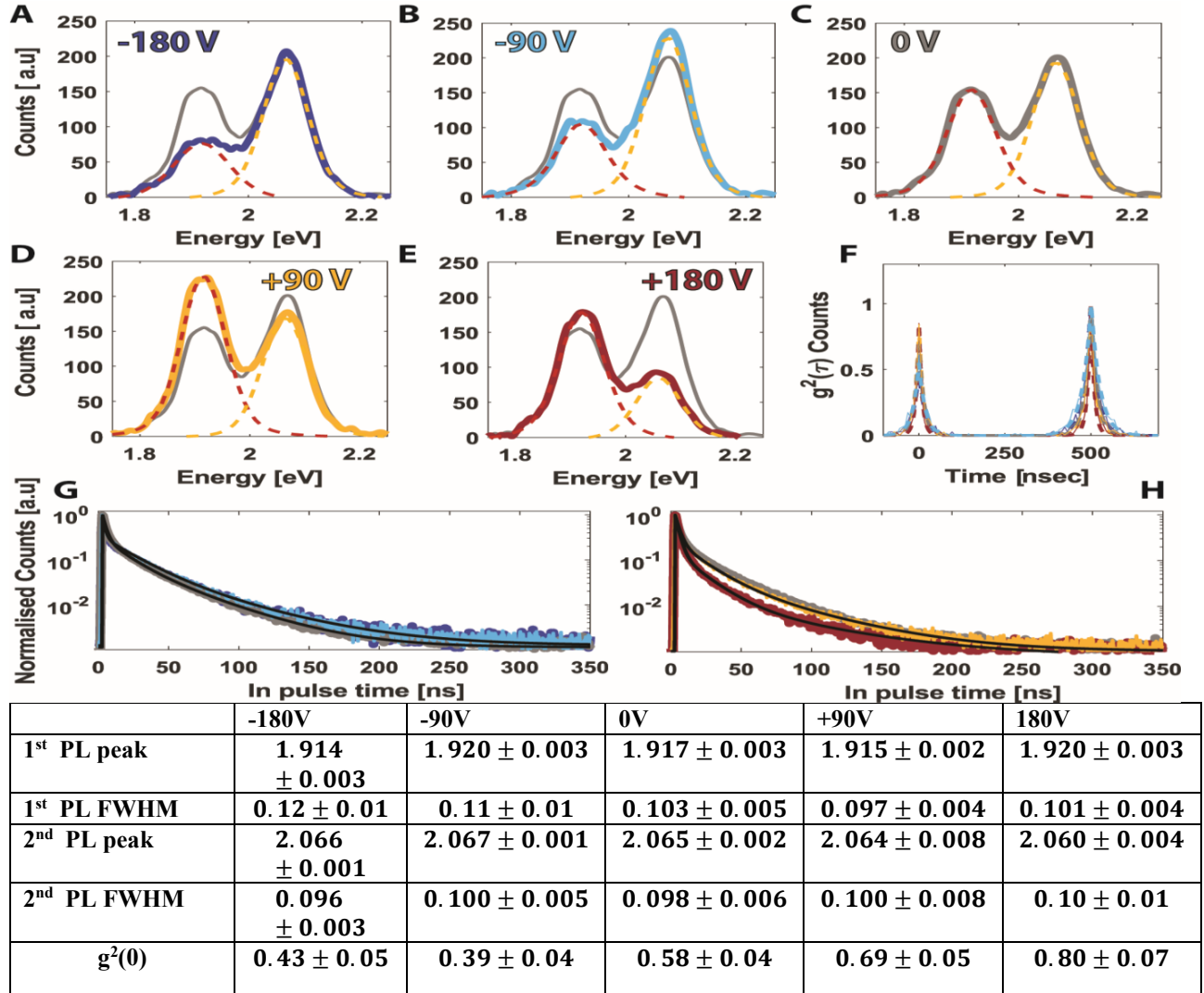

| | -180V | -90V | 0V | +90V | 180V |
|---|---|---|---|---|---|
| **1st PL peak** | 1.914 ± 0.003 | 1.920 ± 0.003 | 1.917 ± 0.003 | 1.915 ± 0.002 | 1.920 ± 0.003 |
| **1st PL FWHM** | 0.12 ± 0.01 | 0.11 ± 0.01 | 0.103 ± 0.005 | 0.097 ± 0.004 | 0.101 ± 0.004 |
| **2nd PL peak** | 2.066 ± 0.001 | 2.067 ± 0.001 | 2.065 ± 0.002 | 2.064 ± 0.008 | 2.060 ± 0.004 |
| **2nd PL FWHM** | 0.096 ± 0.003 | 0.100 ± 0.005 | 0.098 ± 0.006 | 0.100 ± 0.008 | 0.10 ± 0.01 |
| **g²(0)** | 0.43 ± 0.05 | 0.39 ± 0.04 | 0.58 ± 0.04 | 0.69 ± 0.05 | 0.80 ± 0.07 |

**Figure S26. Emission fitting and photon arrival correlations of the "Red-Yellow" Hetero-CQDM in Figure 6.**
(**A-E**) emission spectra of the hetero-CQDM in figure 6A (presumed 2.2 nm/5.3 nm core-shell fused to 1.2 nm/4.8 nm core-shell CdSe/CdS), showing Voigt line fits to each emission center, at different applied voltages between the Aluminum electrodes (3 μm distanced, 900 nm high). **The maximal color switching between the ±90V modulated measurements is 0.152±0.004 eV (48±2nm)** (**F**) Second order correlation function of photons with the g²(0) contrast calculated from the fit, showing lower contrast than the smaller CQDMs presented in figures 2-3 in main text, presumably due to the increased volume which stabilizes the biexciton. (**G, H**) PL life-time traces showing a tri-exponential fit to all EF modulated frame time bins of the measurement.

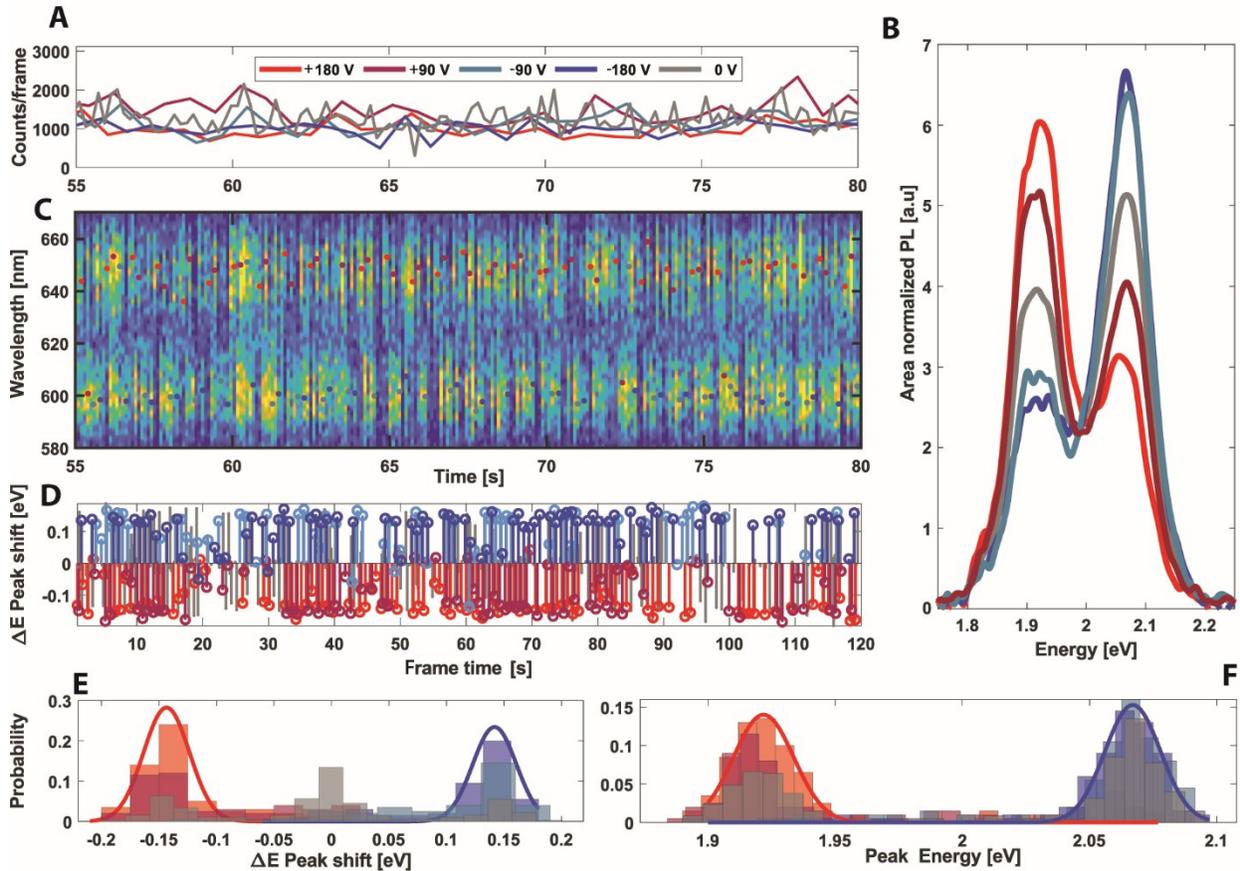

**Figure S27. Emission Intensity and color switching at 10 Hz EF modulation frames of the "Red-Yellow" Hetero-CQDM in Figure 6A.** (**A**) 100 ms binned photon counts on an APD with EF modulation for 25 s of the full measurement, showing relatively stable emission intensity through drastic color switching (**B**) of the accumulated spectra (area normalized). (**C**) frame-by frame emission spectra with colored dots (colors as in the legend in **A**) indicating the gaussian fitted emission peak, showing stable emission at 645 nm with applied +180V and +90V, and alternatively emission at 600 nm when -180V and -90V are applied. (**D**) The energy shift between subsequent frames from the full measurement, shows that for the positive (negative) applied voltages the normalized probable shift (**E**) relative to the next frame is a 150 meV red (blue) shift, due to emission switching between the red-yellow emission centers. The energy shift between frames without applied EF changes randomly between the two cores (gray). (**F**) Single frame emission peak histogram shows the same trend as in (**C**), where the positive and negative applied EF reduce the emission to one side of the CQDM while the emission in 100 ms frames without applied EF is statistically distributed between both centers.

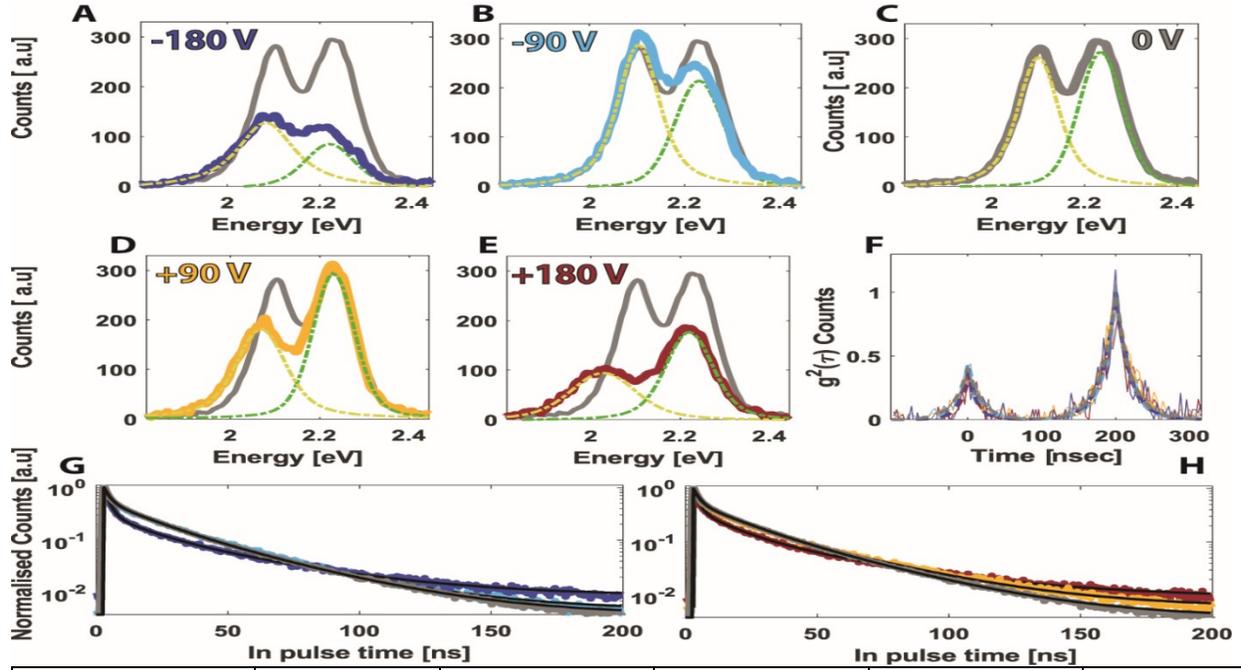

| | -180V | -90V | 0V | +90V | 180V |
|---|---|---|---|---|---|
| **1st PL peak** | 2.084 ± 0.003 | 2.103 ± 0.003 | 2.100 ± 0.003 | 2.068 ± 0.002 | 2.028 ± 0.003 |
| **1st PL FWHM** | 0.15 ± 0.02 | 0.111 ± 0.005 | 0.105 ± 0.005 | 0.134 ± 0.008 | 0.180 ± 0.020 |
| **2nd PL peak** | 2.221 ± 0.004 | 2.230 ± 0.003 | 2.235 ± 0.002 | 2.231 ± 0.002 | 2.220 ± 0.003 |
| **2nd PL FWHM** | 0.130 ± 0.02 | 0.105 ± 0.005 | 0.111 ± 0.005 | 0.111 ± 0.004 | 0.127 ± 0.008 |
| **g²(0)** | 0.25 ± 0.09 | 0.40 ± 0.07 | 0.38 ± 0.04 | 0.28 ± 0.06 | 0.4 ± 0.1 |

**Figure S28. Emission fitting and photon arrival correlations of the "Yellow-Green" Hetero-CQDM in Figure 6. (A-E)** emission spectra of the hetero-CQDM in figure 6C (presumed 1.4 nm/2.8 nm core-shell fused to 1.2 nm/1.8 nm core shell CdSe/CdS), showing Voigt line fits to each emission center, at different applied voltages between the Aluminum electrodes (2.5 µm distanced, 200nm high, with Polyvinylpyrrolidone (PVP)). **The maximal color switching between the ±90V modulated measurements is 0.128±0.004 eV (34±1nm).** The high voltage measurements (±180V) show emission dimming and red shifting of both peaks in addition to the color switching effect, which can indicate a significant reduction of the electron-hole overlap in each of the emission centers. **(F)** Second order correlation function of photons with the g²(0) contrast calculated from the fit, showing higher contrast than the strongly fused CQDMs presented in figures 2-3 in main text, indicating that there can still be improvement in the potential barrier engineering and coupling between the two QDs. **(G, H)** PL life-time traces showing a tri-exponential fit to all EF modulated frame time bins of the measurement. The high voltage measurements (±180V) show a larger amplitude of the "short" life-time component, due to enhanced non-radiative recombination which leads to the dimmed emission intensity and a longer "long" life-time component, in accordance with the assumption of reduced electron-hole wavefunction overlap.

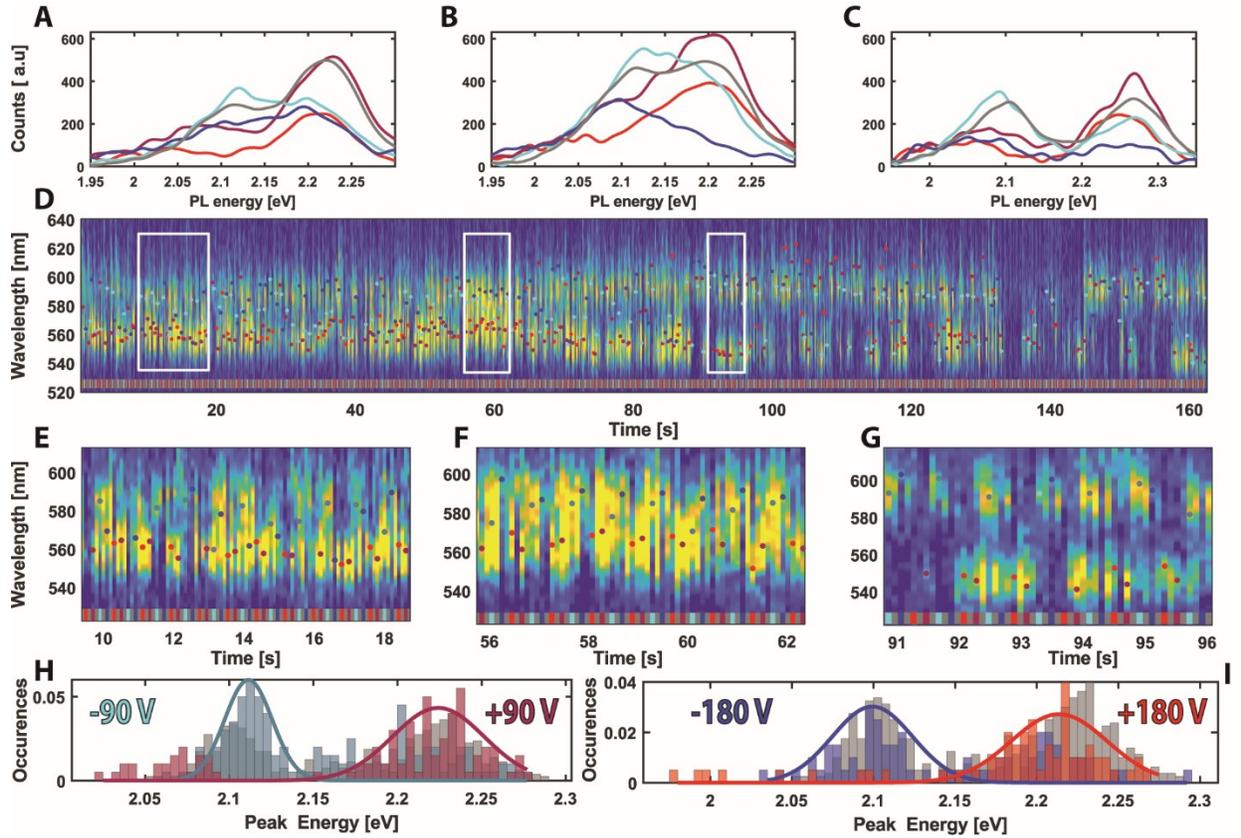

**Figure S29. Color switching at 10 Hz EF modulation frames of the "Yellow-Green" Hetero-CQDM in Figure 6C. (A-C)** Accumulated emission spectra with EF modulation (applied voltages as with colors shown in histograms H-I) of 3 selected time samples from the full measurement (**D**) (in white rectangles), showing that the EF color switching is less stable than for the case of larger CQDM such as presented in figure S26. (**E-G**) the frame-by frame emission spectra with colored dots (colors as in H-I) indicating the gaussian fitted emission peak, showing the emission peak of each frame, overall following the color switching trend as seen from the accumulated spectra. (**H**) the full measurement frame peak histogram for ±90V (colors) and 0V (gray) applied voltage with gaussian fitting to the major emitting peak. It is seen that still there is emission from both peaks under ±90V, with altered population relative to the 0V frames (gray). (**I**) For frames with applied +180V, there is clearly one dominant high intensity peak shown in the distribution, while the -180V frame peak histogram shows less dominant color switching.

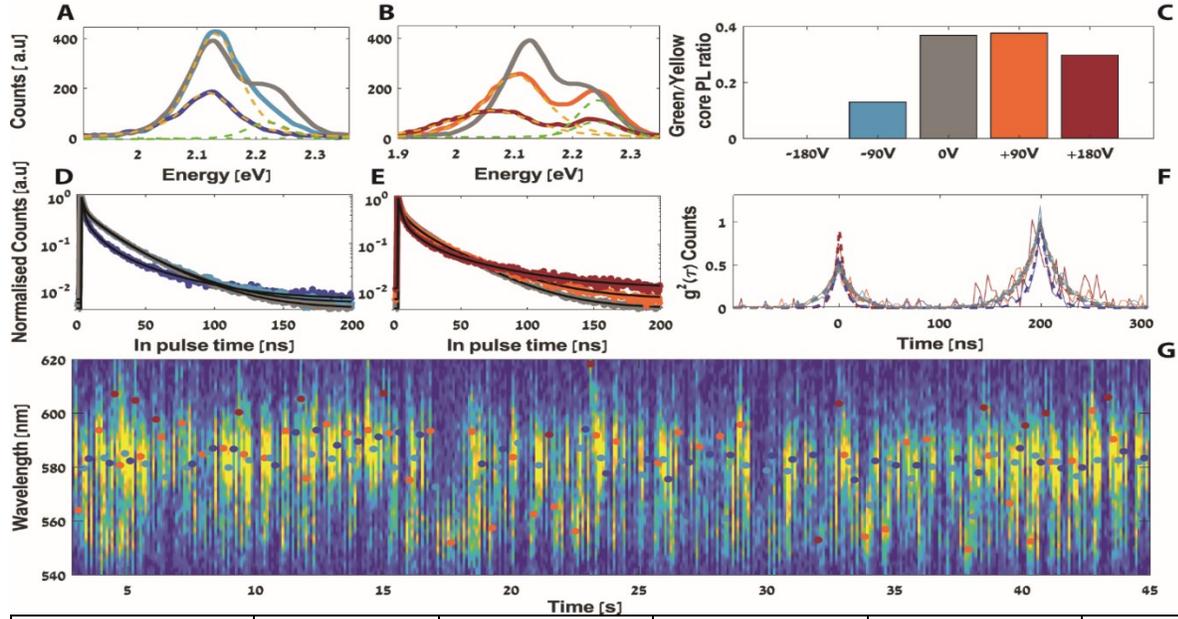

| | -180V | -90V | 0V | +90V | 180V |
|---|---|---|---|---|---|
| **1st PL peak** | 2.120 ± 0.01 | 2.131 ± 0.002 | 2.123 ± 0.002 | 2.102 ± 0.005 | 2.060 ± 0.01 |
| **1st PL FWHM** | 0.15 ± 0.02 | 0.101 ± 0.04 | 0.099 ± 0.003 | 0.130 ± 0.006 | 0.170 ± 0.010 |
| **2nd PL peak** | − − | 2.220 ± 0.002 | 2.229 ± 0.001 | 2.244 ± 0.006 | 2.241 ± 0.003 |
| **2nd PL FWHM** | − − | 0.90 ± 0.01 | 0.093 ± 0.004 | 0.088 ± 0.002 | 0.090 ± 0.01 |
| **g²(0)** | 0.50 ± 0.1 | 0.39 ± 0.07 | 0.40 ± 0.03 | 0.38 ± 0.08 | 0.30 ± 0.2 |

**Figure S30. Partial color switching of a "Yellow-Green" Hetero-CQDM with unequal emission center QY.** (**A**) accumulated spectra for negative and no applied EF with Voigt fitting of the two emission centers (-180V exhibits a single emission center spectra). (**B**) same as **A** but for the opposite polarity of the applied EF. Colors of EF modulated spectra are as shown in the bar graph (**C**), showing the area ratio ($\frac{\int green}{\int yellow}$) of the fitted Voigt peaks from **A, B**. the "Green" core can be switched off completely by a -180V applied EF, while opposite polarity of the field mostly dims the emission from both centers, without changing the population ratio upwards. (**D-E**) PL life-time traces showing a tri-exponential fit to all EF modulated frame time bins of the measurement. The high voltage measurements (±180V) show a similar change relative to the 0 V measurements as explained for the measurement of figure S28. (**F**) Second order correlation function of photons with the g²(0) contrast calculated from the fit. (**G**) 100 ms successive emission spectra of 30 s from the 160 s measurement. The dominant emission peak throughout the measurement is at 584 nm, with EF induced modulation around that wavelength.

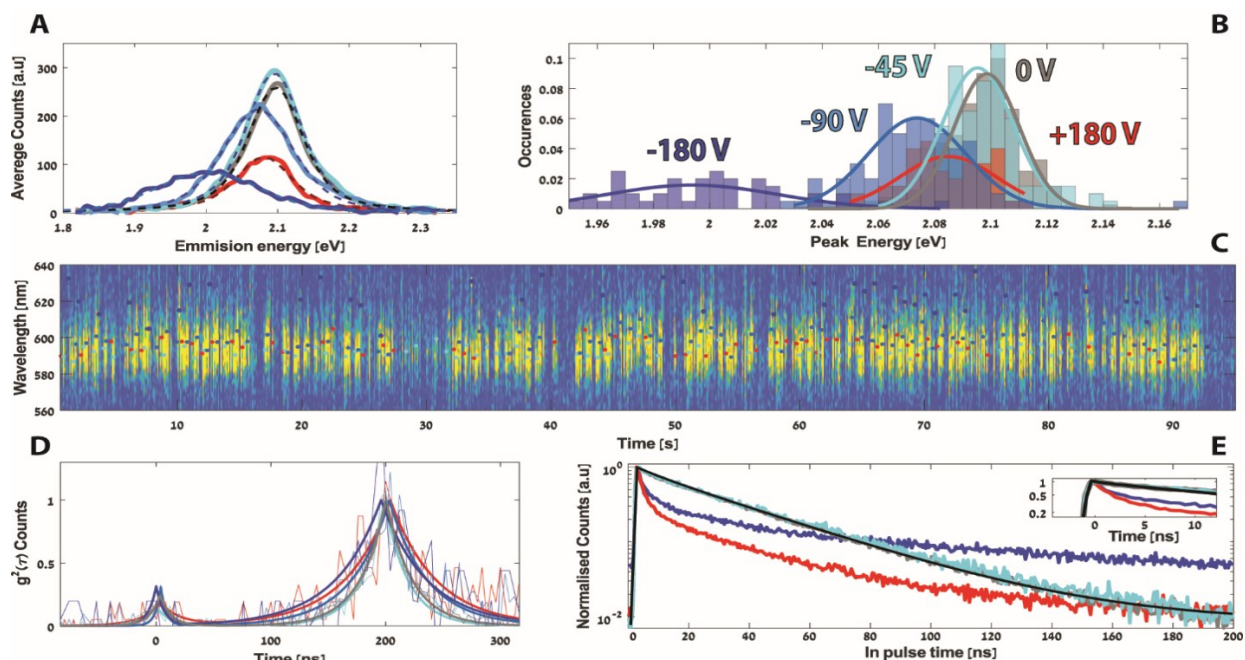

**Figure S31. Intensity switching and spectral broadening of a "Yellow" mono-QD.**

(**A**) Accumulated emission specta at different applied EF (-180V, -90V, -45V, 0V, +180V) with single Voigt fitting for a mono-QD from the same solution as in S25 . The applied EF induces a red shift and emission dimming above 45 V as expected from the quantum confined Stark effect emhancing charge seperation in the QD. This effect can also be seen in (**B**) that presents the single frame emission peak histogram (colors of different voltage data are the same in all panels). An emission shift of 78±5 meV as well as ~70% emission loss between 0V to -180V is observed. This enhanced effect is assumed to be due to choice of PVP as the electrode coating polymer instead of polymethyl methacrylate (PMMA) which drastically reduces the dielectric screening between the QD to the environment and enhances the effective electric field inside the particle. (**C**) 100 ms successive emission spectra of 90 s from the 160 s measurement showing typical mono-QD emission blinking and spectral diffusion. (**D**) Second order correlation function of photons with the $g^2(0)$ contrast calculated from the fit, $g^2(0) < 0.1$ for any applied EF modulation, as expected for CdSe/CdS core/shell QDs of low volume, enhancing non-radiative recombination of biexcitons. (**E**) PL life-time traces showing a multi-exponential life-time of the ±180 V EF modulated emission and a mono-exponential fit for the 0 V, -45 V modulated emission, expected for mono-QDs of this size and structure. The high voltage measurements (±180V) show a similar charge separation affected change relative to the 0V measurements as explained for the measurement of figure S28.

## Supplementary text S1. Self-consistent calculation sequence of the charge carrier states

The wavefunctions and energies of each charge carrier state (hole or electron) are calculated by solving the self-consistent Schrödinger-Poisson equation taking into account the coulomb interaction and self-potential between the two charges, under an external electric field induced potential. A scheme explaining one iteration in the self-consistent calculation is presented below. First the Laplace equation is solved on a 100 nm$^2$ domain with constant potential boundary conditions to simulate the external electric field considering the dielectric screening effects between the NC and the surroundings. $\Psi_{e/h}^n$ and $E_{e/h}^n$ are the electron/hole wavefunctions and energies after n iterations, calculated by the Schrödinger-Poisson equation considering the band potentials ($V_{CB}/V_{VB}$), coulomb interaction of the other charge ($\Phi_e$/ $\Phi_h$) and external electric field potential ($V_E$) without taking into account the self-potential. $\Psi_{e/h}'^{n}$ and $E_{e/h}'^{n}$ are the wavefunctions and state energies calculated taking into account the self-potentials ($\Phi^{SP}_{e/h}$). $\Psi_{e/h}''^{n}$ and $E_{e/h}''^{n}$ are the wavefunctions and state energies calculated taking into account the self-potentials ($\Phi^{SP}_{e/h}$) of the last iteration without the coulomb potential of the other charge carrier. This form of calculation discriminates between the coulomb potential of the other charge carrier and its self-potential. Iterations are done until the energy difference between subsequent iterations converges to less than 1 meV, giving us the wavefunctions and energies of the electron and hole states.

The computational space extends more than 2 times the CQDMs size in each direction with the Dirichlet boundary-condition setting the wavefunction to zero at the edges. We use von Neumann boundary-condition at the inner (between core-shell) and outer boundaries of the QD in order to impose the Ben-Daniel-Duke condition. The applied electric field between the two electrodes is also calculated by solving the Laplace equation on the specific area geometry, considering the dielectric polarization effects of the surroundings (glass, substrate, PMMA, air).

Charge carrier state numerical simulation scheme:

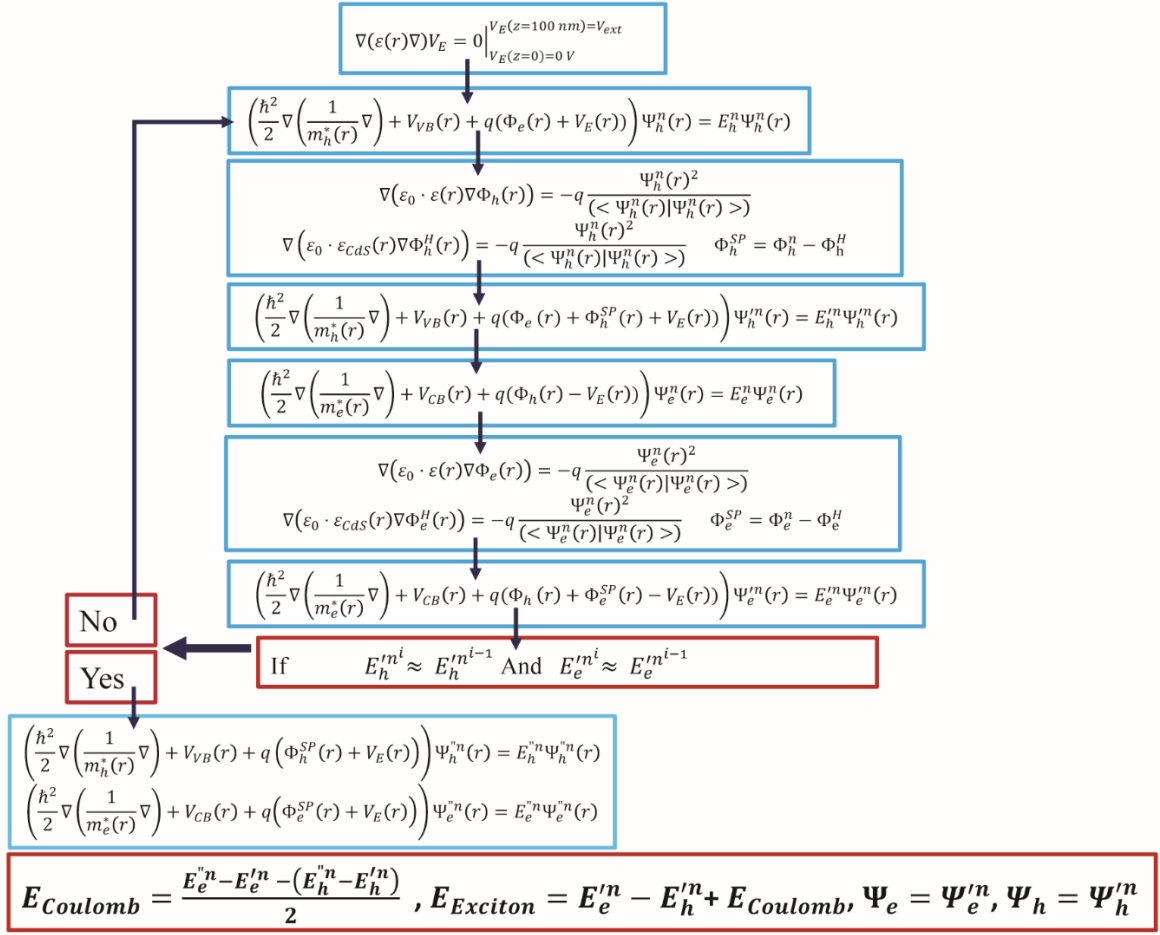

**Table S1.**

Material parameters used in the simulations

|  | CdSe | CdS | Environment | Units | Ref. |
|---|---|---|---|---|---|
| $V_c$ | 1.76 | 1.86 | 5 | [eV] | [7–10] |
| $V_v$ | 0 | -0.64 | -5 | [eV] | [7–10] |
| $m_e^*$ | 0.112 | 0.21 | 1 | $m_0$ | [11] |
| $m_{h\perp}^*$ | 0.48 | 0.376 | 1 | $m_0$ | [11] |
| $m_{h\parallel}^*$ | 1.19 | 0.746 | 1 | $m_0$ | [11] |
| $\varepsilon_\perp$ | 9.29 | 8.28 | 2.4 | - | [11–13] |
| $\varepsilon_\parallel$ | 10.16 | 8.73 | 2.4 | - | [11–13] |

We note that the CdSe/CdS band offset was changed by an additional -0.1 eV to +0.2 eV in the results showing calculations for different band offsets shown in the main article in Fig. 5a[8,14,15]. The dielectric constant of the environment was taken as 2.4, considering the dielectric constant of the solvent-toluene and the coating ligands-oleic acid. The low concentration of poly-methyl-methacrylate (PMMA) in the spin cast QD solution and contradicting reports on its thin film dielectric constant[16,17] being between 2.1-5, lead us to take the lower values for a thin porous matrix of toluene-PMMA-oleic acid.

**Movie S1.**

Step by step lift out and TEM grid preparation (using the DB-FIB SEM) of a 3.5×12 $\mu m^2$ area with measured CQDMs from between the electrodes.

**Movie S2.**

Raw measurement of Photoluminescence emission wavelength on the EMCCD, separated to the Horizontal (top) and Vertical (bottom) polarization components relative to the aluminum electrodes. The 15 second measurement shows the emission from the CQDM presented in figure 1c and 2 in the main article, with periodic EF modulation of +180V to 0V at 10 Hz rate, showing PL emission wavelength switching between subsequent frames due to the EF modulation.

**Movie S3.**

Measurement of Photoluminescence emission wavelength on the EMCCD. The 15 second measurement shows the emission color switching between "red" (~650 nm) and "yellow" (~600 nm) from the CQDM presented in figure 6A in the main article and supplementary figure S27. The Movie shows the 10 Hz rate frames taken with an applied ±180V EF modulation, showing PL emission wavelength switching between subsequent ±180V frames due to the EF modulation.